\documentclass[aps,pra,showpacs,notitlepage]{revtex4}
\usepackage{latexsym}
\usepackage{amsmath}
\usepackage{amssymb}
\usepackage{bm}
\usepackage{wasysym}
\usepackage[dvips]{color}

\usepackage{graphicx}
\usepackage{graphicx}
\usepackage{subfigure}
\usepackage{epsfig}

\newcommand{\hide}[1]{}
\newcommand{\veps}{\varepsilon}

\newcommand{\ra}{\rangle}

\begin{document}

\title{The Landau--Zener Problem with Decay and with Dephasing}

\author{Y. Avishai} 
\affiliation{Department of Physics and the Ilse Katz Center
for Nano-Science, Ben-Gurion University, Beer-Sheva 84105, Israel} 

\author{Y. B. Band}
\affiliation{Department of Chemistry, Department of Physics and
Department of Electro-Optics, and the Ilse Katz Center for
Nano-Science, Ben-Gurion University, Beer-Sheva 84105, Israel}


\begin{abstract}
Two aspects of the classic two-level Landau--Zener (LZ) problem are
considered.  First, we address the LZ problem when one or both levels
decay, i.e., $\veps_j(t) \to \veps_j(t)-i \Gamma_j/2$.  We find that
if the system evolves from an initial time $-T$ to a final time $+T$
such that $|\veps_1(\pm T)-\veps_2(\pm T)|$ is not too large, the LZ
survival probability of a state $| j \ra$ can {\em increase} with
increasing decay rate of the other state $|i \ne j \ra$.  This
surprising result occurs because the decay results in crossing of the
two eigenvalues of the instantaneous non-Hermitian Hamiltonian.  On
the other hand, if $|\veps_1(\pm T)-\veps_2(\pm T)| \to \infty$ as $T
\to \infty$, the probability is {\em independent} of the decay rate.
These results are based on analytic solutions of the time-dependent
Schr\"odinger equations for two cases: (a) the energy levels depend
linearly on time, and (b) the energy levels are bounded and of the
form $\veps_{1,2}(t) = \pm \veps \tanh (t/{\cal T})$.  Second, we
study LZ transitions affected by dephasing by formulating the
Landau--Zener problem with noise in terms of a
Schr\"{o}dinger-Langevin stochastic coupled set of differential
equations.  The LZ survival probability then becomes a random variable
whose probability distribution is shown to behave very differently for
long and short dephasing times. We also discuss the combined effects 
of decay and dephasing on the LZ probability.
\end{abstract}


\maketitle

\section{Introduction} \label{Sec:Intro}

\begin{figure}
\centering\subfigure[]{\includegraphics[width=0.4\textwidth]
{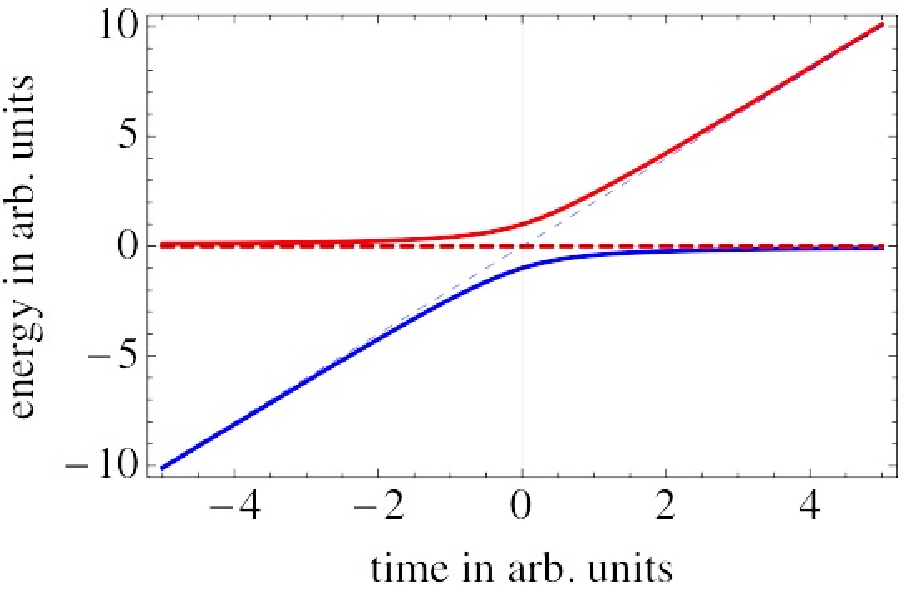}}
\centering\subfigure[]{\includegraphics[width=0.4\textwidth]
{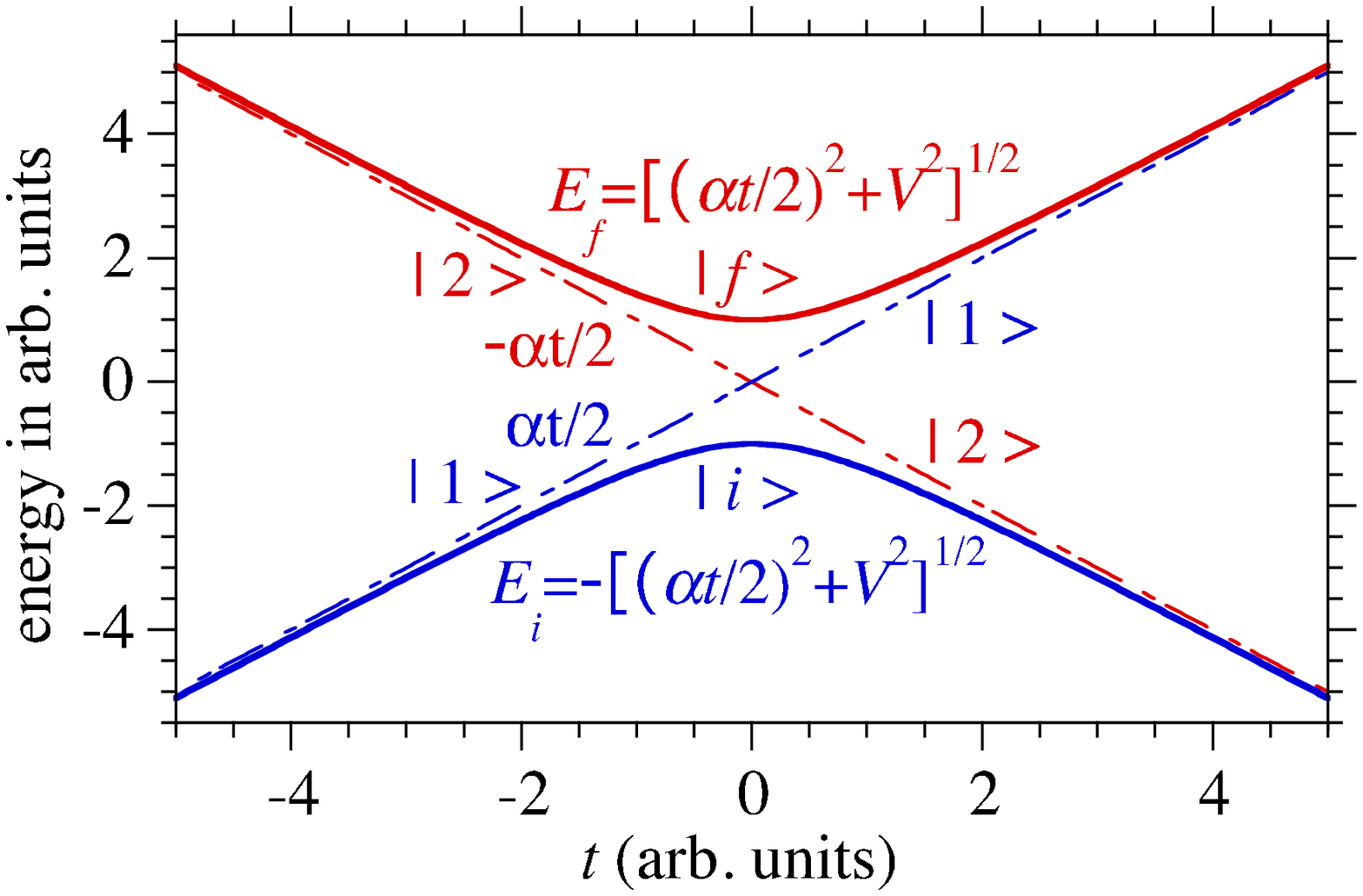}}
\caption{(Color online) (a) Eigenvalues of $H_1(t)$ in Eq.~(\ref{Eq:1}) 
(solid curves) and the diagonal Hamiltonian matrix
elements (dashed lines), with $\alpha = -2$, $V = 1$.  (b) Eigenvalues
of $H_2(t)$ in Eq.~(\ref{Eq:1}) (solid curves) and the diagonal Hamiltonian 
matrix elements (dashed lines), with $\alpha = 2$, $V = 1$.}
\label{Fig_LZ_wo_decay}
\end{figure}

The Landau--Zener (LZ) problem \cite{Landau_32, Zener_32,
Stueckelberg_32, Majorana_32} has been the subject of intense study
for over four score years.  It has become a paradigm time-dependent
two-level dynamical model that has been applied in many areas of
quantum physics.  Here we shall focus on two extensions of this classic
problem.  Let us first briefly remind the reader of the original LZ
problem, since it serves as a starting point for the extensions.  The
LZ problem involves the evolution of the wave function of a coupled
two-level system whose time-dependent energies, when uncoupled, cross
at some time, say $t_0 = 0$ (see Fig.~\ref{Fig_LZ_wo_decay}).  The
relevant physical quantity is the LZ probability, i.e., the modulus
squared of one of the components of the wave function, as time $t \to
\infty$, and it is of interest to determine its dependence on the
coupling strength and on the rate of energy change with time.

In the original version of the problem \cite{Landau_32, Zener_32,
Stueckelberg_32, Majorana_32}, the energy levels depend
linearly on time. Two widely used forms, displayed in 
Figs.~\ref{Fig_LZ_wo_decay}(a) and \ref{Fig_LZ_wo_decay}(b) are, 
\begin{equation}   \label{Eq:1}
    H_1(t) = \left( \! \!
    \begin{array}{cc}
	0 & V  \\
	V & - \alpha t
    \end{array}
    \! \! \right), \quad 
    H_2(t) =  \frac{\alpha t}{2} \sigma_z + V \sigma_x = \left( \!  \!
    \begin{array}{cc}
	\alpha t/2 & V  \\
	V & - \alpha t/2
    \end{array} \! \! \right) .
\end{equation}
Here, $\sigma_{x}, \sigma_{y}$ and $\sigma_{z}$ are the Pauli
matrices, and $\alpha$ is the rate of unperturbed energy change.  A
unitary time-dependent transformation of $H_1(t)$ yields $H_2(t)$.
Figure~\ref{Fig_LZ_wo_decay}(a) plots the eigenvalues of $H_1(t)$ and
Fig.~\ref{Fig_LZ_wo_decay}(b) plots the diagonal elements and the
instantaneous eigenvalues $E_{f,i}(t) = \pm \sqrt{(\alpha t/2)^2 + V^2
}$ of $H_2(t)$ versus time.  As will be argued below, in the presence
of level decay, it is useful to consider other models where, unlike
the linear time-dependent levels, the dependence of the unperturbed
levels $H_{ii}(t)$ on time are such that $H_{ii}(t)$ is bounded at all
times.  One such model is
\begin{equation} \label{th}
    H(t)= \begin{pmatrix} \veps \tanh (t/{\cal T}) & V \\ V & -\veps
    \tanh (t/{\cal T}) \end{pmatrix},
\end{equation}
where $1/{\cal T}$ controls the rate of energy change near $t=0$.

The LZ problem can be stated as follows: 
what is the survival probability $P$ of finding the system in
state $|1 \rangle$ as $t \rightarrow \infty$, if it starts off in the
state $|1 \rangle$ as $t \rightarrow -\infty$ [see
Fig.~\ref{Fig_LZ_wo_decay}(b), where states $|1 \rangle$ and $|2
\rangle$ are the {\em diabatic} states, and $|i \rangle$ and $|f
\rangle$ are the {\em adiabatic} states].  As $\alpha \to 0$, the
adiabatic theorem ensures that the system stays in the initial
adiabatic state $|i \rangle$.  The original LZ problem can be
formulated as follows: Let $P_i(\alpha, V) = \lim_{t \to \infty}
|\psi_i(t)|^2$ denote the survival probability of state $i$ at large
time.  Find $P_i$ and analyze its dependence on $\alpha$ and $V$.

In this paper we consider two extensions of the original problem which 
are of physical interest.  First, we focus on the case where one or both 
of the energy levels decay.   The motivation for studying the effects of 
level decay on the LZ probability is that decay occurs in many physical 
processes, including light-induced transitions between
metastable states \cite{Burshtein_88}, collision-induced losses of
laser-cooled atoms in magneto-optical traps \cite{Marcassa_93,
Band_94}, photoassociative ionization collisions in a magneto-optical
traps \cite{Bagnato_93}, and adiabatic fast passage in nuclear
magnetic resonance population inversion processes in the presence of
radio-frequency magnetic fields {\cite{Hardy_86}, to mention a few.

Level decay can be modeled by letting the diagonal elements of the
Hamiltonian acquire a time-independent negative imaginary part,
${\mathrm{Im}}[H_{ii}] = - \Gamma_i/2$, where $\Gamma_i/\hbar$ is the
decay rate of level $i, \ (i=1,2)$.  For example, consider the case where
$H_1(t)$, Eq.~(\ref{Eq:1}), is modified such that the matrix element
$H_{22} = -\alpha t-i \Gamma/2$.  Transitions between decaying
states were analyzed both using a master equation approach and by 
adding a decay term to the Hamiltonian in Ref.~\cite{Burshtein_88}.
Here we obtain analytic solutions using the latter method and analyze 
our results in terms of avoided level crossing in the complex energy 
plane.

In the absence of coupling, $V=0$ in 
Eq.~(\ref{Eq:1}), the corresponding diabatic wave function propagates as 
$\psi_2(t) = \exp[(-i (\alpha t- \alpha T - i \Gamma)(t+T)/2] \psi_2(-T)$.  
But with $V \ne 0$, the decay of level 2 affects the survival probability 
of level 1 in a non-trivial way, and we obtain the {\it Landau--Zener 
problem with  decay}.  The central goal of this problem is to determine 
the probability $P(T)=|\psi_1(T)|^2$ that the adiabatic state $\psi_1(T)$ 
is occupied in the far future, given the initial condition that in the far 
past it was fully occupied, $|\psi_1(-T)|^2 = 1$.  The reason for insisting 
on a finite (albeit large) time $T$ will become evident below. 

For the Hamiltonian $H_1(t)$ in Eq.~(\ref{Eq:1}), modified by adding 
${\mathrm{Im}}[H_{22}] = - \Gamma /2$, the LZ problem with decay was 
addressed by Akulin and Schleich~\cite{Akulin_Schleich_92}.  The probability 
$P(\infty)$, is found to be independent of $\Gamma$ \cite{comment_AS}.
We show below that this result is due to the divergence of $|H_{11}(t) -
H_{22}(t)|$ as $|t| \to \infty$, i.e., it is in some sense an artifact. 
One of our main goals is then to study models for the LZ problem with decay 
where $|H_{11}(t) - H_{22}(t)|$ is bounded as $|t| \to \infty$, and show 
that in this case, $P(T)$ does depend on $\Gamma$.
Work on related problems has been reported in Refs.~\cite{Band_94,
Burshtein_88, Dridi_10, Dridi_12, Scala_11, Uzdin_12, Kocharovsky_77,
Kocharovsky_rev, Suominen_98, Vitanov_97, Moyer_01, Benderskii_04,
Graefe_06, Schilling_06, Shore_06, Fainberg_07, Castro_08},
Refs.~\cite{Gefen_87, Rammer, Saito_07} consider a LZ transition for
two states coupled to a bath of harmonic oscillators, and
Refs.~\cite{Rammer, Saito_07} find that at zero temperature there is
no influence of the environment on the transition probability, in a
fashion similar to the Akulin and Schleich
result~\cite{Akulin_Schleich_92}.

The second extension considered here concerns the case where the LZ 
transition is affected by dephasing.  Dephasing of a quantum system 
occurs due to interaction between the system and its
environment.  Examples include collisions of a particle with other
particles, and interactions with environmental degrees of freedom such
as an electromagnetic field that is random or stochastic.  In the case
of dephasing due to collisions with particles, each collision can have
a random duration and a random strength; in the case of interactions
with an environment, the many degrees of freedom of the environment
(the ``bath'') can randomly affect the phase of the wave function.
This results in a time-dependent uncertainty $\delta(\varphi(t))$ in
the phase of the wave function.  At a time $t=\tau$ for which
$\delta(\varphi(\tau)) = 2 \pi$, interference is completely lost.
Incorporation of dephasing in LZ transitions has been extensively
studied \cite{Rammer, Efrat1, Efrat2, Pokrovsky, Avron}.  Dephasing
processes occur in metals \cite{Aleiner_01}.  Moreover, dephasing is
important in atomic clock transitions \cite{ChangYeLukin_04}, in
quantum information processes \cite{qc} and in nuclear-spin-dependent
ground-state dephasing of diamond nitrogen-vacancy centers \cite{NV}.
We treat such transitions using a Schr\"{o}dinger-Langevin stochastic
differential equation formalism \cite{vanKampenBook}, and solve the
time-dependent Schr\"odinger equation with a Gaussian white-noise
stochastic term, and with Ornstein--Uhlenbeck noise.  This enables us
to study not only the averaged survival probability but also its
distribution and its dependence on the strength of coupling between
the system and the environment.  As we shall see, the distribution in
the strong coupling regime (short dephasing time) is very different
from that in the weak coupling regime, both for white-noise and for
Ornstein--Uhlenbeck noise.

The outline of the paper is as follows.  Section~\ref{Sec:TDSE}
formulates the LZ problem with decay.  In Sec.~\ref{IIA} the 
problem is cast as a set of two uncoupled second order differential equations.  
This formalism is used to arrive at analytic solutions in 
Sec.~\ref{SubSec:Eq_1} for the time-dependent Schr\"odinger 
equations derived from $H_1(t)$ in Eq.~(\ref{Eq:1}) and 
in Sec.~\ref{SubSec:tanh} for the Hamiltonian in  Eq.~(\ref{th}), properly 
modified to include decay terms.  Section~\ref{Sec:Num_Examples} presents 
numerical and analytical examples worked out with these Hamilonians.
Section~\ref{SubSec:decay_both_levels} describes the dynamics of LZ
when both levels decay.  Section~\ref{Sec:dephasing} considers LZ
transitions with dephasing due to interaction with an environment.
Finally, Sec.~\ref{Sec:Summary} contains a summary and conclusion.

\section{The Landau--Zener problem with decay}  \label{Sec:TDSE}

In this section we formulate and solve the LZ problem with decay.
The approach is to replace the set of two coupled first order 
differential equations by a set of two uncoupled second order 
differential equations.  Analytic solutions of the time-dependent 
Schr\"odinger equations are obtained for $H_1(t)$ in 
Eq.~(\ref{Eq:1}), and for the Hamiltonian of Eq.~(\ref{th}), with the 
diagonal elements modified to have a time-independent negative imaginary 
part. The solutions of the resulting differential equations are obtained
in terms of transcendental functions and expressions for the wave 
functions that satisfy the appropriate boundary conditions are presented.

\subsection{Derivation of second order differential equations for 
$\psi_i(t)$}  \label{IIA}

The most general form of the LZ problem is encoded in the
time-dependent Schr\"dingier equation for the two-component spinor
$\psi=\binom{\psi_1(t)}{\psi_2(t)}$ that include also initial
condition at time $t=-T$ for large $T$,
\begin{equation} \label{LZgeneral1}
    i \dot {\psi}=H \psi = \begin{pmatrix} Z_1(t)&V\\V&Z_2(t)
    \end{pmatrix}, \quad \psi_1(-T)=1, \ \ \psi_2(-T)=0.
\end{equation}
Expressing $\psi_2(t)$ in terms of $\psi_1(t)$ by using the first
equation, and substituting into the second equation we find,
\begin{equation} \label{LZgeneral2}
    \ddot{\psi}_1(t)+i[Z_1(t)+Z_2(t)]
    \dot{\psi}_1(t)+[V^2-Z_1(t)Z_2(t) + i \dot{Z}_1(t)]\psi_1(t)=0,
    \quad \psi_1(-T)=1, \ \ \dot{\psi}_1(-T)=-iZ_1(-T)~.
\end{equation} 
Both Hamiltonians in Eq.~(\ref{Eq:1}) can be written in the form of
Eq.~(\ref{LZgeneral1}), and a diagonal time-dependent transformation can 
transform from one form to the other.

\subsection{Solution of the Akulin-Schleich Version}
\label{SubSec:Eq_1}

Consider the Hamiltonian $H_1(t)$ in Eq.~(\ref{Eq:1}), modified to include 
an imaginary part in $H_{22}(t)$,
\begin{equation} \label{ASD}
H=\begin{pmatrix} 0 & V \\ V &  -\tfrac{1}{2}(\alpha t + i \Gamma) 
\end{pmatrix},
\end{equation}  
where for convenience we replace $\alpha \to \alpha/2$.  Here $\alpha>0$ ($[\alpha]=$ energy/time), $V > 0$ and $\Gamma >0$
are constants ($[V]=[\Gamma] =$ energy).  It is useful to define a
dimensionless time $\tau$, a dimensionless adiabaticity parameter
$\lambda$, and a dimensionless decay parameter $\beta$.  Restoring
$\hbar$, these are defined as,
\begin{equation}  \label{dimless}
    t \to \xi \tau \ \ (\xi \equiv \sqrt{\tfrac{\hbar} {2
    \alpha}} , \ \ [\xi]=\mbox{time}), \quad \lambda \equiv
    \frac{V}{\sqrt { \alpha \hbar}}, \quad \beta = \frac{\Gamma}{
    \sqrt{ \alpha \hbar}}.
\end{equation}
Renaming the dimensionless time to be $t$, instead of $\tau$, we
obtain the dimensionless version of the
Hamiltonian used in Ref.~\cite{Akulin_Schleich_92} is
\begin{equation}  \label{Eq:Ht'_dim}
    {\cal H}(t) =\left( \!  \!  \begin{array}{cc} 0 & \ \ \  \lambda \\
    \lambda & \ \ -\tfrac{1}{2}(t + i \beta) \end{array} \!  \!  \right) \equiv
     \left( \!  \!  \begin{array}{cc} 0 & \lambda \\
    \lambda &z(t) \end{array} \!  \!  \right) .
\end{equation}
This is a special case of the $2 \times 2$ Hamiltonian defined in 
Eq.~(\ref{LZgeneral1}) with $Z_1(t)=0$, and $Z_2(t) = z(t) \equiv -
\tfrac{1}{2}(t + i \beta)$ and $V \to \lambda$.   
The time-dependent Schr\"{o}dinger
equations take the form,
\begin{subequations}  \label{SL}
\begin{equation}  \label{SL1}
    i {\dot \psi}_{1}(t)=\lambda \psi_{2}(t),
\end{equation}
\begin{equation}  \label{SL2}
    i {\dot \psi}_{2}(t) = \lambda \psi_{1}(t) + z(t) \psi_{2}(t) .
\end{equation}
\end{subequations}
Employing the procedure detailed in arriving Eq.~(\ref{LZgeneral2}), the
second-order differential equations for $\psi_1(t)$ and $\psi_2(t)$ are,
\begin{subequations} \label{Soln_AS}
\begin{equation} \label{SL6} 
    \ddot{\psi}_1-\tfrac{1}{2}(i t -\beta) \dot{\psi}_1 + \lambda^2
    \psi_1 = 0 ,
\end{equation}
\begin{equation} \label{SL3}
    \ddot{\psi}_2 -\tfrac{1}{2}(i t -\beta) \dot{\psi}_2 +
    (\lambda^2-\tfrac{i}{2}) \psi_2 = 0 ,
\end{equation}
\end{subequations}
with the initial conditions for $\psi_1(t)$ being,
\begin{equation} \label{ICAS}
    \psi_1(-T) = 1, \quad \dot{\psi}_1(-T) = -i \lambda \psi_2(-T)=0 .
\end{equation} 
The most general solution of each second order differential equation
is a linear combination of two basic solutions of the differential
equations Eq.~(\ref{SL6}) and Eq.~(\ref{SL3}) given respectively as,
\begin{subequations}  \label{BasicAS}
\begin{equation} \label{BasicAS1}
F_{11}(t) = D \left(-2 i \lambda^2,\tfrac{1}{2}(e^{i \frac{\pi
}{4}}t+e^{i \frac{3 \pi}{4}}\beta) \right ), \quad F_{12}(t) = M
\left(i \lambda^2, \tfrac{1}{2}, [\tfrac{1}{2} (e^{i \frac{\pi }{4}}t 
+ e^{i \frac{3 \pi}{4}}\beta)]^2 \right).
\end{equation}
\begin{equation} \label{BasicAS2}
F_{21}(t) = D \left(-1-2 i \lambda^2, \tfrac{1}{2}(e^{\frac{3 i
\pi}{4}} \beta+e^{\frac{ i \pi}{4}}t) \right) , \quad F_{22}(t) =
(e^{\frac{3 i \pi}{4}}\beta + e^{\frac{ i \pi}{4}}t)M \left(1+i
\lambda^2, \tfrac{3}{2},(e^{\frac{3 i \pi}{4}}\beta+e^{\frac{ i
\pi}{4}}t)^2 \right) .
\end{equation}
\end{subequations}
Here $D(a,z)$ is the parabolic cylinder function of order $a$ and
argument $z$ \cite{AS_65} while $M(a,b,z)$ is the regular Kummer
(confluent Hypergeometric) function \cite{AS_65}.  Both are entire
functions of $z$.  The first index of the subscripts refers to the
function $\psi_1$ or $\psi_2$ while the second refers to the
appropriate term in a linear combination defining the functions (see
below). Thus we have,
\begin{equation} \label{psibc}
\psi_i(t)=C_{i 1}F_{i 1}(t)+C_{i 2}F_{i 2}(t), \ \ i=1,2~.
\end{equation} 
Using these solutions and the initial conditions (\ref{ICAS}), we can
obtain expressions for the coefficients $C_{11}$ and $C_{12}$ and
$\psi_1(T)$ at any time $T$.  (Practically, instead of taking $T \to
\infty$ we choose large but finite $T$ such that the survival depends
on the decay rate $\Gamma/\hbar$).
Denoting the Wronskian of the two basic solutions by $\Delta
\equiv F_{11}(-T)\dot{F}_{12}(-T)-F_{12}(-T) \dot{F}_{11}(-T)$, the
coefficients are given by
\begin{equation} \label{solC}
    C_{11} = [\dot{F}_{12}(-T)+\tfrac{T}{2 i} \dot{F}_{11}(-T)] /
    \Delta, \quad C_{12}=-[F_{12}(-T)+\tfrac{T}{2 i} F_{11}(-T)] /
    \Delta .
\end{equation}
Using Eq.~(\ref{psibc}), we finally obtain an expression for the wave
function $\psi_1(T)$,
\begin{equation} \label{psi1T}
    \psi_1(T) = \frac{1}{\Delta} \{ [\dot {F}_{12}(-T)+\tfrac{T}{2 i}
    \dot{F}_{11}(-T)] F_{11}(T)-[F_{12}(-T) + \tfrac{T}{2 i}
    F_{11}(-T)]F_{12}(T) \} .
\end{equation}
This expression can be used directly to calculate
$P(T;\lambda,\beta)$.  Accurate results require high precision
evaluation of the parabolic cylinder and confluent hypergeometric
functions for large and complex argument and parameters.
Alternatively, we can solve the differential equations numerically.
The results will be discussed in Sec.~\ref{SubSec:AS_numerical}.

At this point we can understand why, in Ref.~\cite{Akulin_Schleich_92}, 
the survival probability turned out to be independent 
of the decay rate $\beta$. The reason is that in the linear case, 
the dependence on the decay constant $\beta$ enters only
through the {\it argument} of the transcendental functions, not
through the {\it parameters}. Explicitly, the corresponding
arguments are $(t +i \beta/2)$ and $[(1+i)t-(1-i)\beta/2)]$. In the 
limit $T \to \infty$, their dependence on $\beta$ is minuscule.  The 
coefficients $C_{ij}$ are determined through the
initial conditions at $-T \to -\infty$ whereas the probability
$P(T;\lambda,\beta)$ is calculated at large $T \to \infty$.  In both
cases $\beta$ can be neglected as $T \to \infty$.  Hence, we arrive at
the conclusion that, {\it if the diagonal energies diverge as $T \to
\pm \infty$, the probabilities are independent of $\beta$}.  
Hence, this result is an artifact of the divergence of the diabatic 
and adiabatic energies and because $\beta$ enters the solution only 
through the arguments of the transcendental functions. To alleviate this 
problem, one may require cutting off the linear divergence at some large 
but finite $T$, such that $\veps_i(|t|>T)=\veps_i(T)$. This choice will 
be employed in Sec.~\ref{SubSec:AS_numerical}. Alternatively, one might 
use another version of the LZ Hamiltonian where $H_{ii}(t)$ are bounded 
for all times. This choice is explained in Sec.~\ref{SubSec:tanh} and
employed in Sec.~\ref{SubSec:psiTh}.

\subsection{Solution for the case $H_{11}=\veps \tanh (t/{\cal T})$ and 
$H_{22}=-\veps \tanh (t/{\cal T}) - i \tfrac {\Gamma}{2}$} 
\label{SubSec:tanh}

Instead of using energies that depend linearly on time and doing the
propagation from $\pm T$, here we consider energies that depend smoothly 
on time and saturate beyond a time ${\cal T}$.  Specifically, we
consider the Hamiltonian
\begin{equation} \label{Htanh}
    H(t)= \begin{pmatrix} \veps \tanh (t/{\cal T}) & V \\ V & -\veps
    \tanh (t/{\cal T}) - i \tfrac {\Gamma}{2} \end{pmatrix},
\end{equation}
where the dimensions of the quantities appearing in the Hamiltonian
are $[\veps]=[V]=[\Gamma]=$ energy, $\veps$ determines the saturation
energy, $V$ is the strength of coupling and $\veps/ {\cal T}$ is the
slope of the energy curve at $t=0$.  Defining dimensionless time and
energies, we have
\begin{equation} \label{Htanh_dim_var}
    \tau = t/{\cal T} \to t, \quad \chi = \veps {\cal T}/\hbar,
    \quad \lambda = V {\cal T}/\hbar, \quad \beta=\Gamma {\cal T}/(2
    \hbar).
\end{equation}
Scaling the Hamiltonian such that $\chi=1$, the dimensionless Hamiltonian 
becomes
\begin{equation} \label{THDimensionless}
    {\cal H} = \begin{pmatrix} \tanh t & \lambda \\ \lambda & -(\tanh
    t + i \beta) \end{pmatrix},
\end{equation}
and $t$, $\lambda$ and $\beta$ are dimensionless.  In addition to being a 
realistic form that can be experimentally realized, the advantage of 
choosing this parametrization leads to an analytic solution for the wave 
function that acquire a relatively simple form as $t \to \infty$. In this 
expression, the dependence of the survival probability on the decay rate 
$\beta$ is more transparent.  

The coupled time-dependent Schro\"dinger equations are,
\begin{equation} \label{SEtanh}
    i \dot {\psi}_1(t) = \tanh t \, \psi_1(t)+\lambda \psi_2(t), \quad i
    \dot {\psi}_2(t) = -(\tanh t+i \beta) \psi_2(t) + \lambda
    \psi_1(t) .
\end{equation}
Straightforward manipulations lead to second-order equations for
$\psi_{1,2}(t)$, of which we will concentrate on that for $\psi_1(t)$
that has a general expression as in Eq.~(\ref{psibc}).  The
initial conditions are,
\begin{equation} \label{BCth}
    \psi_1(-\infty) = 1, \quad \psi_2(-\infty) = 0, \quad
    \dot{\psi}_1(-\infty) = i, \quad \dot{\psi}_2(-\infty) = -i
    \lambda .
\end{equation}

The functions $F_{11}(t)$ and $F_{12}(t)$ are rather complicated; they
have the general form, 
\begin{equation} \label{Fk}
F_{1k}(t; \lambda,\beta)=f_k(e^{2t};\lambda,\beta) F \left(
a_k(\lambda,\beta),b_k(\lambda,\beta) ,c_k(\lambda,\beta);
\frac{e^{2t}}{1+e^{2t}} \right) , \ \ k=1,2,
\end{equation}
in which $f_k(x; \lambda,\beta)$ is an {\it algebraic function} of
$x=e^{2t}$ and $F(a,b,c;y)$ is the Hypergeometric function
\cite{AS_65}.  The parameters of the Hypergeometric functions are
algebraic functions of $\lambda, \beta$, but they will not be
specified here since we give below the closed-form of $\psi_1(T)$ for
large (but finite) $T$.  Thus, unlike the former case, the dependence
of $\psi_1(T)$ on the decay rate $\beta$ enters not through the
argument of the transcendental functions but through its parameters
$a_k$, $b_k$, and $c_k$, where $k=1,2$.  Moreover, the Hypergeometric
functions are required only at the endpoints.  This is especially
convenient because $ -\infty < t < \infty$, which implies $0 \le
y=e^{2t}/(1+e^{2t}) \le1$.  In practice, the argument
$y=e^{2t}/(1+e^{2t})$ virtually reaches the limits $(0,1)$ for
$(-T,T)=(-10,10)$ where $F$ and $\dot{F}$ are simply given by
\cite{AS_65}:
$$F(a,b,c;0)=1, 
\quad F(a,b,c;1) = \frac{\Gamma(c) \Gamma(c-a-b)}
{\Gamma(c-a)\Gamma(c-b)} \equiv \Lambda(a,b,c),
$$ 
\begin{equation} \label{Hyper}
 \frac{d F(a,b,c;y)}{dy}=\frac{ab}{c}F(a+1,b+1,c+1;y) , \quad 
\dot{F}(a,b,c;y) = \frac{d F(a,b,c;y)}{dy} \frac{-2e^{2t}}{(1+e^{2t})^2} 
\to 0. 
\end{equation}
The analogous equation of (\ref{psi1T}) is,
\begin{equation} \label{psi1Th}
    \psi_1(T)=\frac{1}{\Delta(-T)} \{ [\dot{F}_{12}(-T) - i
    \dot{F}_{11}(-T)]F_{11}(T)-[F_{12}(-T)-iF_{11}(-T)]F_{12}(T) \} .
\end{equation}
The advantage of the present approach is as follows.  Using the
abbreviation $f_k(e^{\pm T}) \to f_k(\pm T)$ and the definitions
(\ref{Fk}) of $F_{1k}(t)$ combined with the properties of the
Hypergeometric
functions specified in Eqs.~(\ref{Hyper}) we have,
$$\dot{F}_{12}(-T) = \dot{f}_2(-T)F(a_2,b_2,c_2;0)+f_1(-T)
\dot{F}(a_2,b_2,c_2;0) = \dot{f}_2(-T). $$
$$ \dot{F}_{11}(-T) = \dot{f}_1(-T). \quad F_{11}(T)=f_1(T)  
\Lambda(a_1,b_1,c_1).$$
$$ F_{12}(-T)=f_2(-T) , \quad F_{11}(-T)=f_1(-T). \quad
F_{12}(T) = f_2(T)\Lambda(a_2,b_2,c_2).$$ 
$$ \Delta(-T) = f_1(-T)\dot{f_2}(-T)-\dot{f_1}(-T)f_2(-T).$$
Substitution into Eq.~(\ref{psi1Th}) yields,
\begin{equation} \label{psi1f}
    \psi_1(T) = \frac{[\dot{f}_2(-T)-i \dot{f}_1(-T)]f_1(T) 
    \Lambda(a_1,b_1,c_1)-[f_2(-T)-i f_1(-T)]f_2(T) \Lambda(a_2,b_2,c_2)}
    {f_1(-T)\dot{f_2}(-T)-\dot{f_1}(-T)f_2(-T)} .
\end{equation}
The algebraic functions $f_1(e^{2t})$ and $f_2(e^{2t})$ are known
explicitly but will not be specified here, because we directly
present the closed-form expression for $\psi_1(T)$ employing the
replacements $1+e^{2 T} \to e^{2T}$, and $1+e^{-2 T} \to 1$.  Defining
the quantities,
$$ s_\pm \equiv \sqrt{(\beta \pm 2 i)^2-4 \lambda^2},$$
the result is,
\begin{equation} \label{psiTh}
    \psi_1(T) = \frac{1}{s_+-\beta}e^{-[\beta+ \tfrac{1}{2}(s_+ +
    s_-)]T} \Gamma(\tfrac{1}{2}s_+)
\end{equation}
$$ \left\{ \left[ \frac{e^{ s_+ T}(s_+-\beta-2i)\Gamma(1-\tfrac{s_+}{2})}
{\Gamma[\tfrac{1}{4}(s_--s_+-4 i)]\Gamma[\tfrac{1}{4}(s_--s_++4+4 i)]} \right] 
 + \left [ \frac {[2i e^{2 s_+T} 
 +e^{\tfrac{1}{2}i \pi s_++s_+T} (\beta-s_+)] \Gamma
 (1+\tfrac{s_+}{2})}{\Gamma[\tfrac{1}{4}(s_-+s_+-4 i)]
 \Gamma[\tfrac{1}{4}(s_-+s_++4+4 i)]}
  \right ]  \right \}.$$
It should be pointed out that $\psi_1(T)$ decays to zero as $T \to \infty$
because, for $T$ large enough such that $\tanh T \approx 1$, the vector
$\psi \equiv \binom {\psi_1}{\psi_2}$ propagates with the constant 
Hamiltonian ${\cal H} \approx \binom {1 \ \ \ \ \ \ \ \lambda} {\lambda \ \ 
-(1+i \beta)}$, and the corresponding evolution operator 
$\exp{(-i {\cal H} T)}$ vanishes as $T \to \infty$ when $\beta > 0$.

\section{Numerical Results for the LZ problem with one decaying level}
\label{Sec:Num_Examples}

In this section we present numerical results for the LZ problem with 
decay using the Hamiltonians specified in Eq.~(\ref{Eq:Ht'_dim}) and in 
Eq.~(\ref{THDimensionless}).  In the first case we solve the 
pertinent differential equation numerically, 
and focus an the probability $P(t;\lambda,\beta)=|\psi_1(t)|^2$ as a 
function of time.  In the second case we use the analytic expression 
(\ref{psiTh}) that is true at large time $|t|>T$ (where 
$\tanh (T/{\cal T}) \simeq 1$.  Physical aspects to be explored are: 
(1) St\"uckleberg oscillations as function of time.
(2) St\"uckleberg oscillations as function of coupling strength $\lambda$ 
and decay rate $\beta$.
(3) Non-monotonic behavior of $P(T;\lambda,\beta)$ as function of $\beta$. 
Sec.~\ref{SubSec:decay_both_levels} shows results for the case where both 
levels decay.

\subsection{Results based on the Hamiltonian in Eq.~(\ref{Eq:Ht'_dim})}
\label{SubSec:AS_numerical}

We first discuss the results for the linear case (without saturation)
defined by Eq.~(\ref{Eq:Ht'_dim}).  The analytical expression of
$\psi_1(T)$ can be formally obtained by substitution of the solutions
in Eq.~(\ref{BasicAS}) into expression (\ref{psi1T}).  However, we
find it instructive to inspect the probability $P(t;\lambda,\beta) =
|\psi_1(t)|^2$ at all times, despite the fact that the LZ problem
focuses on the probability at infinite time.  For that reason we
prefer to numerically integrate Eqs.~(\ref{SL1}) and (\ref{SL2}) with
initial conditions $\psi(-T) = \binom{1} {0}$, and thereby obtain the
two-component wave function $\psi(t) = \binom {\psi_1(t)}{\psi_2(t)}$
for specific values of the parameters $\lambda$ and $\beta$.  \\

\noindent 
\underline {\bf Behavior of $|\psi_i(t)|^2$ for $-T \le t \le T$}: \\
The time-dependent probabilities $|\psi_i(t)|^2$ for $i = 1$ and 2 are
plotted versus time $-T<t<T$ in Fig.~\ref{LZ_w_decay_beta_10_l_0.3}(a)
for $T=40, \lambda = 0.3$ and $\beta = 10$, and in
Fig.~\ref{LZ_w_decay_beta_10_l_0.3}(b) for $ T=10, \lambda = 0.3$ and
$\beta = 10$.  For comparison, the results without decay ($\lambda =
0.3$ and $\beta = 0$) are plotted as dashed curves.  The main features
observed in Fig.~\ref{LZ_w_decay_beta_10_l_0.3}(a) are: (1) Rapid
St\"uckelberg oscillations for $\beta=0$ whose amplitudes diminish
with time.  (2) Still for $\beta=0$, the St\"uckelberg oscillations of
$|\psi_i(t)|^2$ saturate at large times $T$ and approach the
prediction of the decay free LZ formula.  (3) For $\beta = 10$, the
population of the diabatic state 2 (red solid curve) stays close to
zero throughout the whole time interval [see inset of
Fig.~\ref{LZ_w_decay_beta_10_l_0.3}(a)].  (4) The population of the
diabatic state 1 (blue solid curve) saturates at a value $P(T;
\lambda, \beta)$ that is slightly higher than $P(T; \lambda, 0)$.  In
Fig.~\ref{LZ_w_decay_beta_10_l_0.3}(b), where $T=10 (< 40)$ the
St\"uckelberg oscillations with time for $\beta=0$ are still
significant at $t=T$ and the value of$P(T; \lambda, 0)$ is much higher
than for $T=40$.  This confirms our statement that for smaller $T$,
the sensitivity to decay is more significant.  The reason for the
inequality $\Delta P \equiv P(T; \lambda,\beta) > P(T; \lambda,0)$
will be explained below.\\

\begin{figure}
\centering\subfigure[]{\includegraphics[width=0.4\textwidth]
{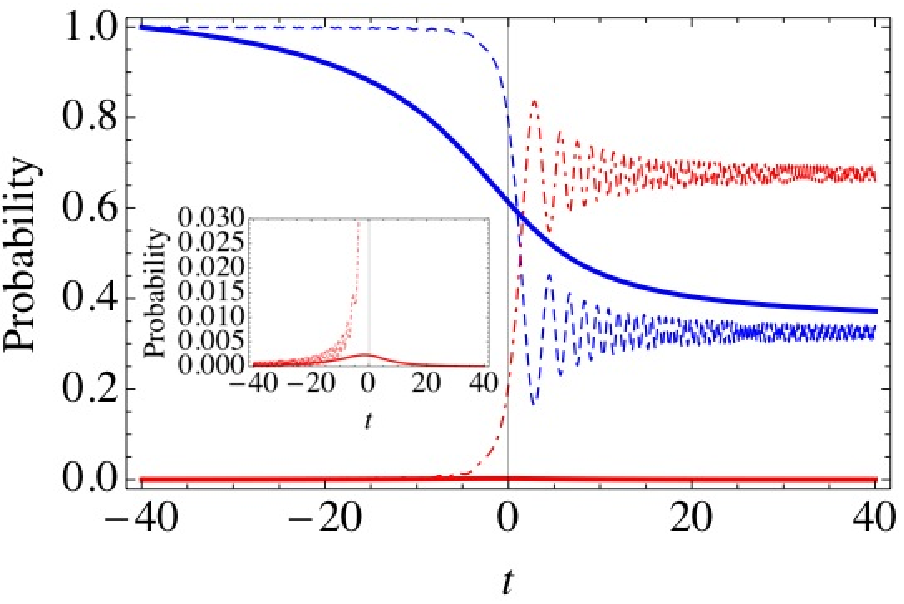}}
\centering\subfigure[]{\includegraphics[width=0.4\textwidth]
{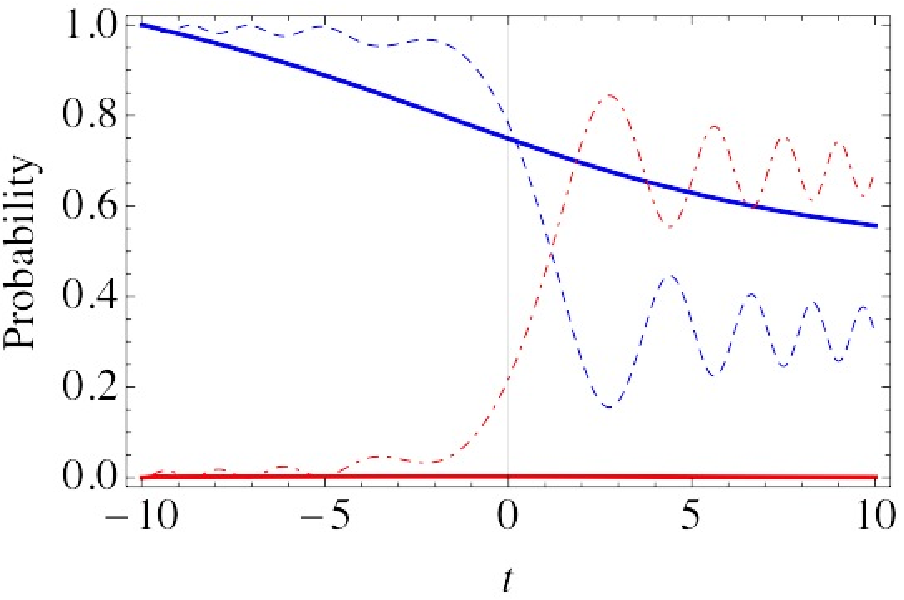}}
\caption{(Color online) (a) Probabilities $|\psi_i(t)|^2$ for $i = 1$
and 2 versus time for $T=40$, $\lambda = 0.3$ and $\beta = 10$.  For
comparison, the dashed curves are without decay ($\beta = 0$).  The
inset is a blowup at very small probability.  (b) Same as (a), except
$T = 10$.}
\label{LZ_w_decay_beta_10_l_0.3}
\end{figure}

Let us now turn to the unexpected result that for large $\beta$,
$\Delta P \equiv P(T; \lambda,\beta)-P(T; \lambda,0)>0$ as evident in
Figs.~\ref{LZ_w_decay_beta_10_l_0.3}.  We already stressed that this 
occurs at finite $T$.  Moreover, we see from 
Fig.~\ref{LZ_w_decay_beta_10_l_0.3} that $P(T=10)>P(T=40)$.  Upon
taking the limit $T \to \infty$, we find $\Delta P (T) \to 0$ in
accordance with the result in Ref.~\cite{Akulin_Schleich_92}.

To understand how decay can increase the survival probability
$P(T;\lambda,\beta)$, it is instructive to consider the eigenvalues of
the Hamiltonian in Eq.~(\ref{Eq:Ht'_dim}) as a function of time in
a fashion similar to Ref.~\cite{VZ_11} where the influence of level 
widths on anti-crossing was discussed.  Inspection of the complex 
eigenvalues yields the following condition for the crossing of the real 
part of the eigenvalues at $t=0$ \cite{VZ_11} (see also 
Sec.~\ref{SubSubSec:eig}),
\begin{equation} \label{cross_cond}
    \beta \ge 4 \lambda .
\end{equation}
This is shown in Fig.~\ref{Fig_LZ_E_1}(a) which plots the real part of
the eigenvalues of the Hamiltonian in Eq.~(\ref{Eq:1}) versus time and
in Fig.~\ref{Fig_LZ_E_1}(b) which shows the imaginary parts for
$\lambda=1$.  For $\beta > 4$, the real parts of the two eigenvalues
cross at $t=0$ while the imaginary parts do not.  On the other hand,
for $\beta < 4 \lambda$ crossing is avoided.
Figure~\ref{Fig_LZ_E_1}(c) and (d) are similar to
Fig.~\ref{Fig_LZ_E_1}(a) and (b) but for a smaller decay rate, $\beta
= 3$, where there is an avoided crossing (as opposed to a crossing).
[Similarly, for $\beta > 1.2$ the real part of the eigenvalues cross
(and the imaginary part of the eigenvalues do not) when $\lambda =
0.3$ (as used in Fig.~\ref{LZ_w_decay_beta_10_l_0.3}).  We chose to
plot the $\lambda = 1$ results in Figs.~\ref{Fig_LZ_E_1}(a) and (b)
and and in (c) and (d) because it is easier to see the results when
the curves are farther apart.]  As the decay rate $\beta$ increases
beyond a critical value, the real part of the eigenvalues cross,
rather than undergoing an avoided crossing as is the case for $\beta =
0$.  Hence, the probability at the final time, $P(T;\lambda,\beta)$,
increases with increasing $\beta$ for sufficiently large $\beta$.

\begin{figure}
\centering\subfigure[]{\includegraphics[width=0.4\textwidth]
{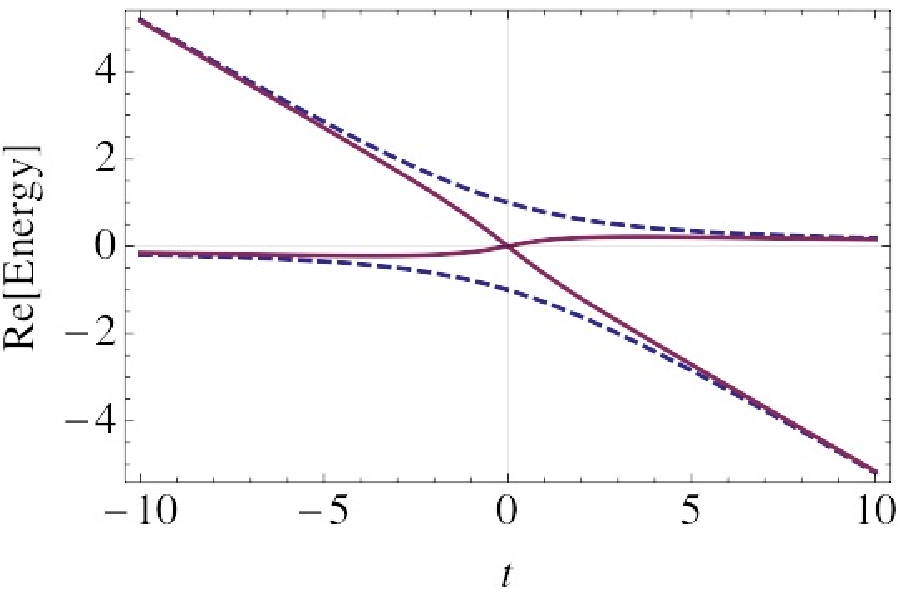}}
\centering\subfigure[]{\includegraphics[width=0.4\textwidth]
{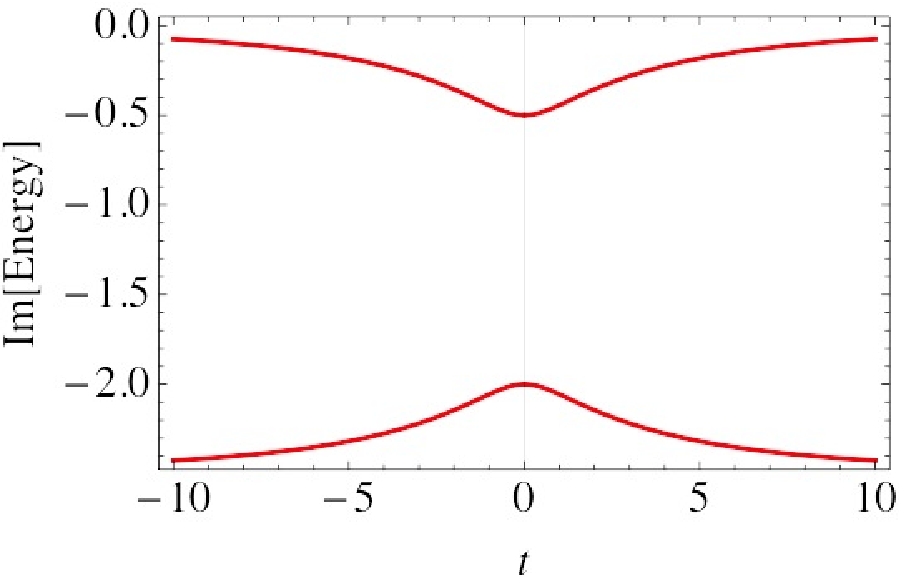}}
\centering\subfigure[]{\includegraphics[width=0.4\textwidth]
{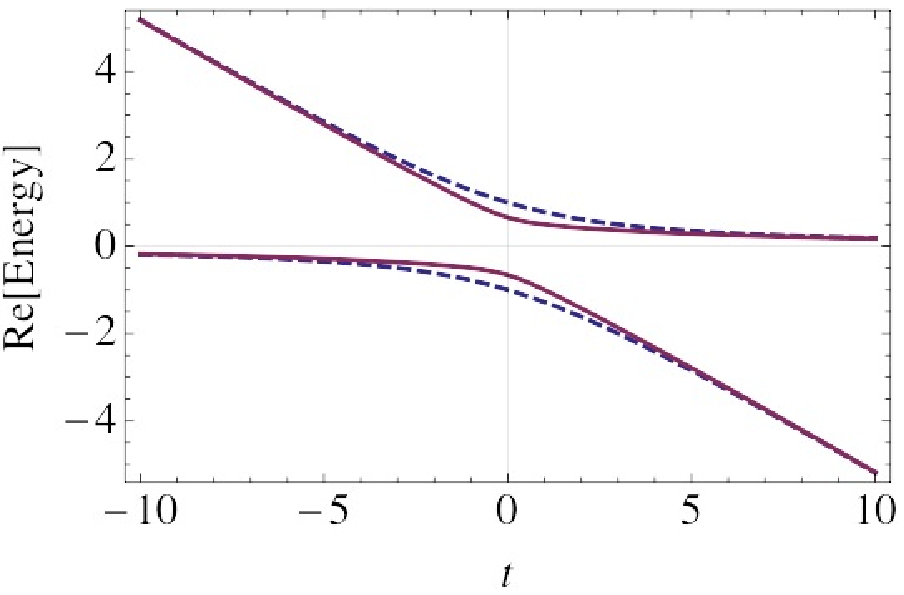}}
\centering\subfigure[]{\includegraphics[width=0.4\textwidth]
{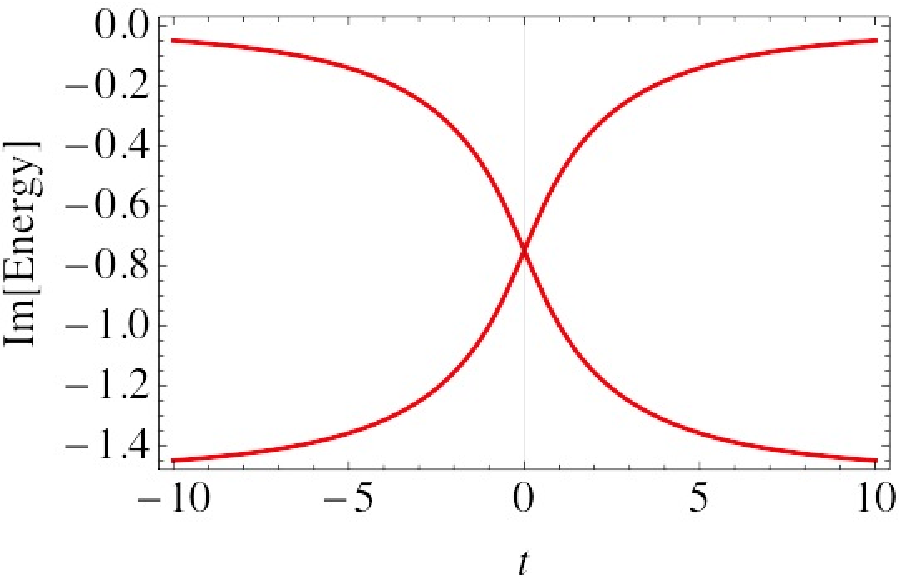}} 
\caption{(Color online) (a) Real part of the eigenvalues of the
Hamiltonian in Eq.~(\ref{Eq:1}) with $\lambda = 1$, $\beta = 5$ as a
function of time (solid red curve).  For comparison, the blue dashed 
curves are the results without decay ($\lambda = 1$, $\beta = 0$), and 
have an avoided crossing.  (b) Imaginary part of the eigenvalues with 
$\lambda = 1$, $\beta = 5$.
(c) Real part of the eigenvalues of the Hamiltonian in
Eq.~(\ref{Eq:1}) with $\lambda = 1$ and $\beta = 3$ as a function of
time  (solid red curve).  For comparison, the blue dashed curves are for 
$\lambda = 1$ and $\beta = 0$, and have a {\em larger} splitting.  
(d) Imaginary part of the eigenvalues with $\lambda = 1$ and $\beta = 3$.}
\label{Fig_LZ_E_1}
\end{figure}


We now plot the probability $P(T;\lambda,\beta)$ as a function of
$\lambda$ and $\beta$.  Figure~\ref{Fig_Prob_lambda_beta}(a) shows the
results for $T = 40$ and Figure~\ref{Fig_Prob_lambda_beta}(b) is for
$T = 10$.  The general trend is that the probabilities decrease
significantly with increased $\lambda$ and also increase with
increasing $\beta$, but the increase with $\beta$ is much more
significant for~\ref{Fig_Prob_lambda_beta}(b).  Near $\beta = 0$, the
first of a series of recurring oscillation peaks that arise from the
St\"{u}ckelberg oscillations evident in
Fig.~\ref{LZ_w_decay_beta_10_l_0.3} is evident in part (b).  This
series of peaks stretches on as a function of $\lambda$ at small
$\beta$ but is not visible because the figure only goes up to $\lambda
= 10$.  These oscillatory peaks are much smaller in magnitude and are
farther appart in $\lambda$ for Fig.~\ref{Fig_Prob_lambda_beta}(a)
which is for $T = 40$.

\begin{figure}
\centering\subfigure[]{\includegraphics[width=0.45\textwidth]
{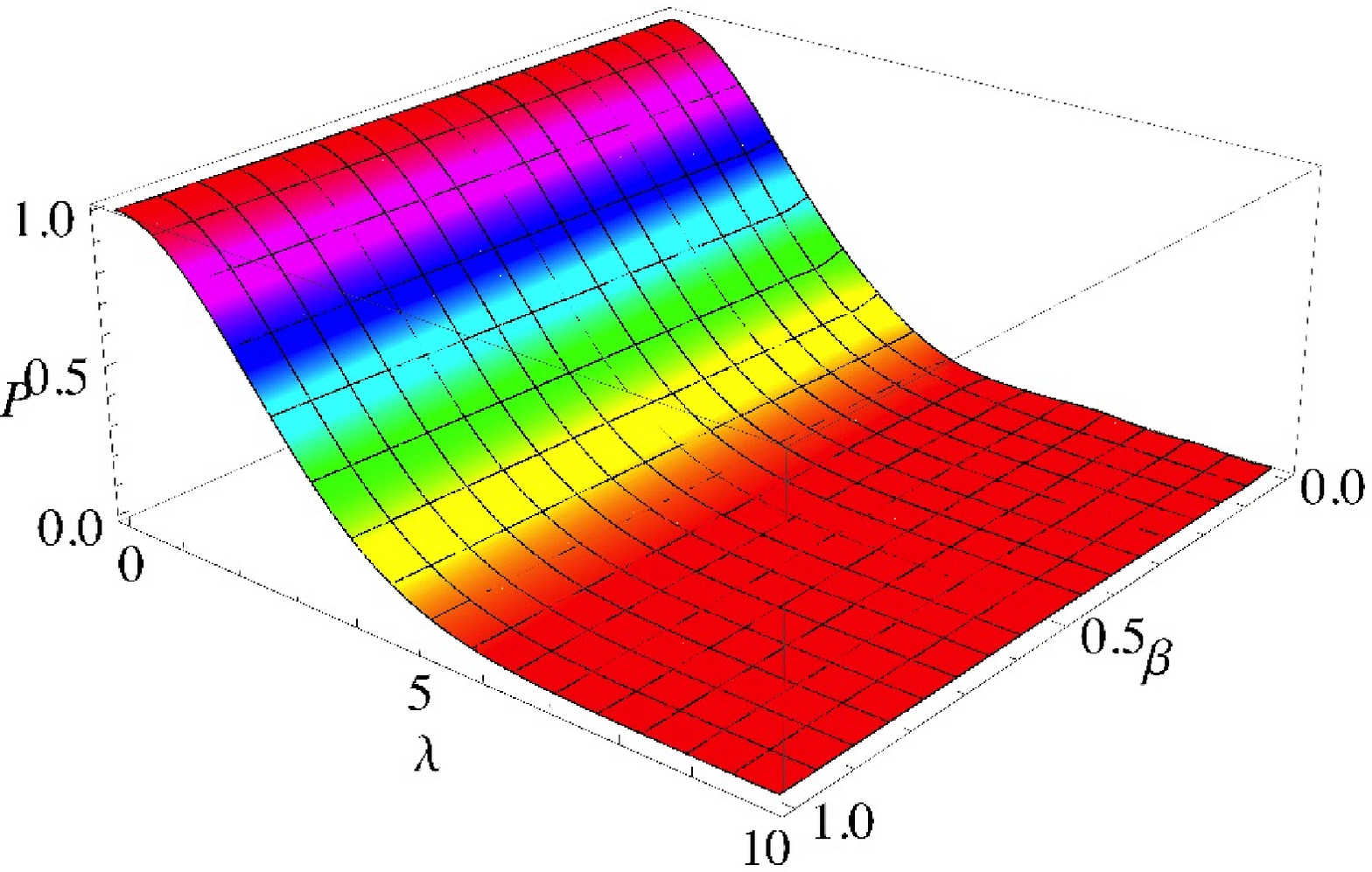}}
\centering\subfigure[]{\includegraphics[width=0.45\textwidth]
{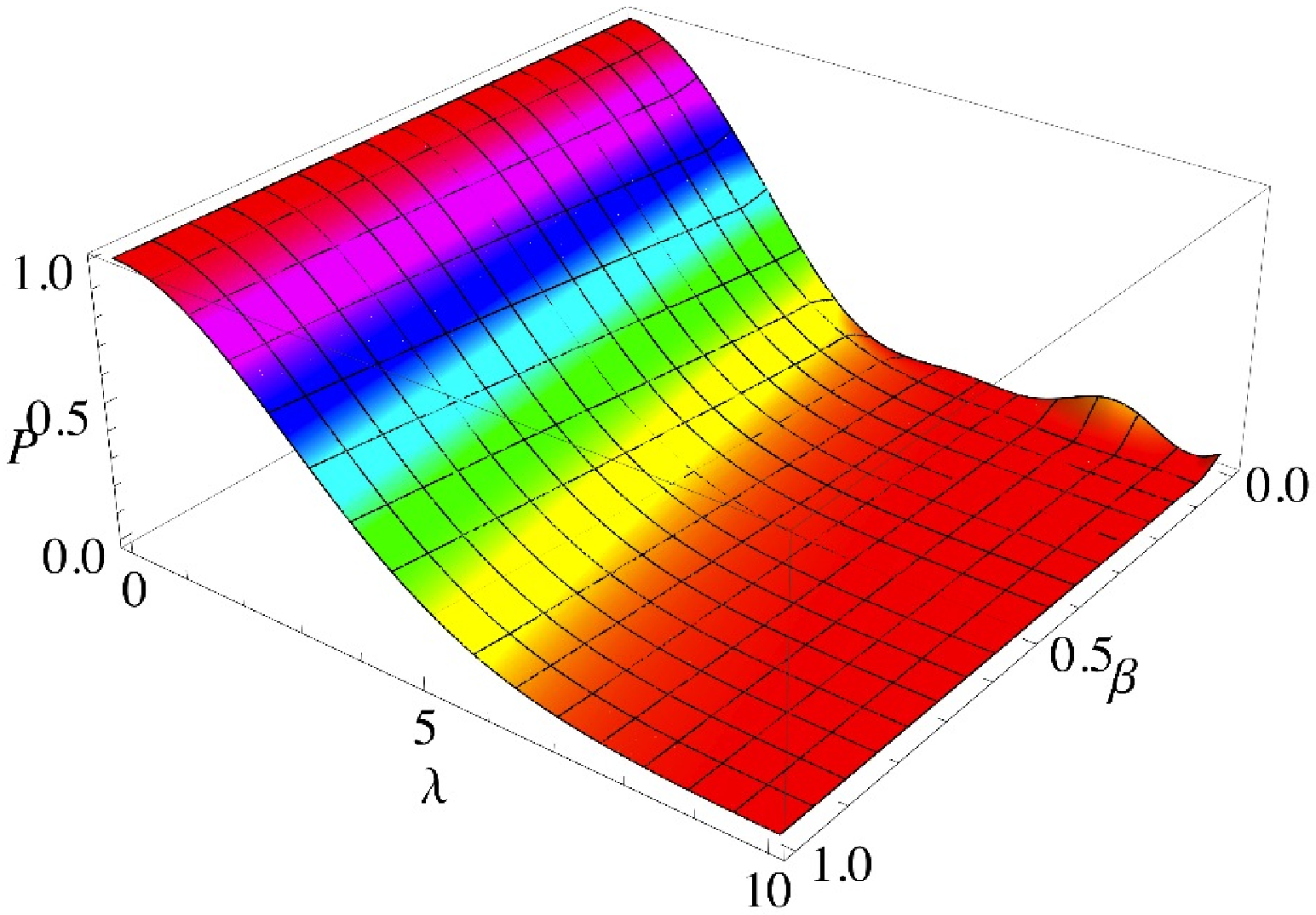}} \caption{(Color online) (a)
Probability $P(T;\lambda,\beta)$ versus $\lambda$ and $\beta$ for
$T=40$.  (b) Probability $P(T;\lambda,\beta)$ versus $\lambda$ and
$\beta$ for $T=10$.}
\label{Fig_Prob_lambda_beta}
\end{figure}



\subsection{Results based on the Hamiltonian (\ref{THDimensionless}) 
(saturated energies)}
\label{SubSec:psiTh}

In this section we present results for the saturated of energy levels
as specified in the Hamiltonian of Eq.~(\ref{THDimensionless}).  The
results are qualitatively similar to those presented previously but
the analytic expression (\ref{psiTh}) enables a simpler and more
transparent analysis.  Unlike the previous discussion we will focus
here only on the long time behavior, beyond which the levels are
virtually saturated.  First we carry out an elementary analysis of the
eigenvalues and find the same condition, $\beta \ge 4 \lambda$, for
level crossing as in Eq.~(\ref{cross_cond}).  Then we use
(\ref{psiTh}) to analyze the behavior of $P(T;\lambda,\beta)$.  Our
analysis includes first a study of the St\"uckleberg oscillations as 
a function of the coupling strength $\lambda$ for large $T$, and
second, the study of situations where the survival probability  {\it 
increases} as $\beta$ approaches (and then surpasses) $4 \lambda$ 
from below.

\subsubsection{Analysis of the eigenvalues}  \label{SubSubSec:eig}

In the original LZ problem with linear time dependence of the diagonal
elements of the Hamiltonian, without decay, the probability depends
crucially on how close the adiabatic energy levels are to one another.
Specifically, for small $\lambda$, $P(\infty)$ is high, and for large
$\lambda$, $P(\infty)$ decays as $e^{-C/\lambda^2}$ where $C$ is a
constant.  But what happens if there is a decay term, where the
Hamiltonian is not hermitian and its eigenvalues are complex?  To
answer this question it is useful to investigate the instantaneous
eigenvalues as a function of time.  The eigenvalues of the Hamiltonian
$\binom{\tanh t \ \ \ \ \ \ \ \ \ \ \ \ \lambda}{\lambda \ \ \ \ \
-(\tanh t+i \beta}$ are,
\begin{equation} \label{eigenTH}
    \veps_{1,2}(t) = \frac{1}{8}\left[ -2 i \beta \pm
    \frac{\sqrt{2}}{\cosh t} \sqrt{8 i \beta \sinh t-\{
    16(1-\lambda^2)+\beta^2 +[\beta^2-16(1+\lambda^2)]\cosh 2t \}
    }\right] .
\end{equation}
Crossing (complex) levels occurs when $\veps_1=\veps_2$, namely, the
expression inside the square root should vanish for some value(s) of
$t$.  Closer inspection shows that a real solution can occur only for
$t=0$, where the expression inside the square root equals $2(16
\lambda^2-\beta^2)$.  From this simple analysis we can draw the
following conclusions [results (1)-(4) pertain
to the case $4 \lambda \ge \beta$ while result (5) pertains to $4
\lambda < \beta$] \cite{linear_eig}:\\
(1) The square root at $t=0$ is real, therefore
${\mathrm{Im}}[\veps_1(0)$] = ${\mathrm{Im}}[\veps_2(0)] = -\beta/4$,
i.e., the imaginary parts of the (complex) energies cross at $t=0$.\\
(2) Im[$\veps_i(t)$] is an antisymmetric function of $t$ with respect
to the crossing point -$\beta/4$.\\
(3) Re[$\veps_i(t)] \ne 0 $ is a symmetric function of $t$, and
${\mathrm{Re}}[\veps_1(t)] = - {\mathrm{Re}}[\veps_2(t)]>0$.  Hence,
the real parts of the energies do not cross for $\beta < 4 \lambda$.\\
(4) As $\beta \to 4 \lambda$ from below, Re[$\veps_1(0)] \to 0^+$ and
Re[$\veps_1(0)] \to 0^-$.  Combined with result (1), the complex 
energy eigenvalues cross at $t=0$.  \\
Points (1)-(4) are summarized in Figs.~\ref{EV1}(a) and (b).\\
(5) For $4 \lambda > \beta$ the real parts of the complex energies
cross at $t=0$ but the imaginary parts do not.  The probability
depends mainly on the behavior of the real parts of the eigenvalues,
so that $P(T;\lambda,\beta)$ increases with $\beta$ even in this case.
Result (5) is summarized in Figs.~\ref{EV1}(c) and (d).
Hence, we conclude that for large $\beta$, the probability
$P(T;\lambda,\beta)$ is a slowly {\it increasing function} of $\beta$.
This somewhat counter intuitive result is confirmed by our analytical
result (\ref{psiTh}) for $\psi_1(T)$.
\begin{figure}[!ht]
\centering\subfigure[]{\includegraphics[width=0.24\textwidth]{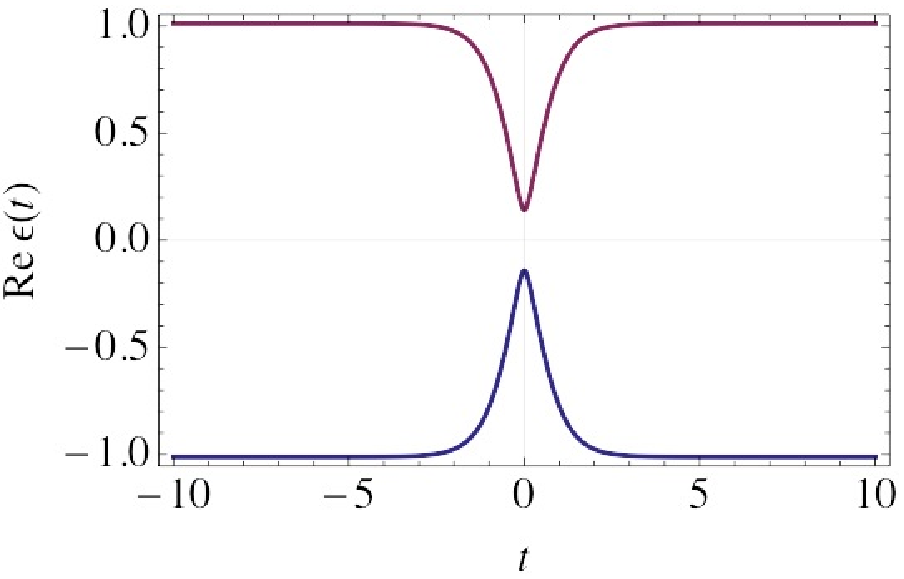}}
\centering\subfigure[]{\includegraphics[width=0.24\textwidth]{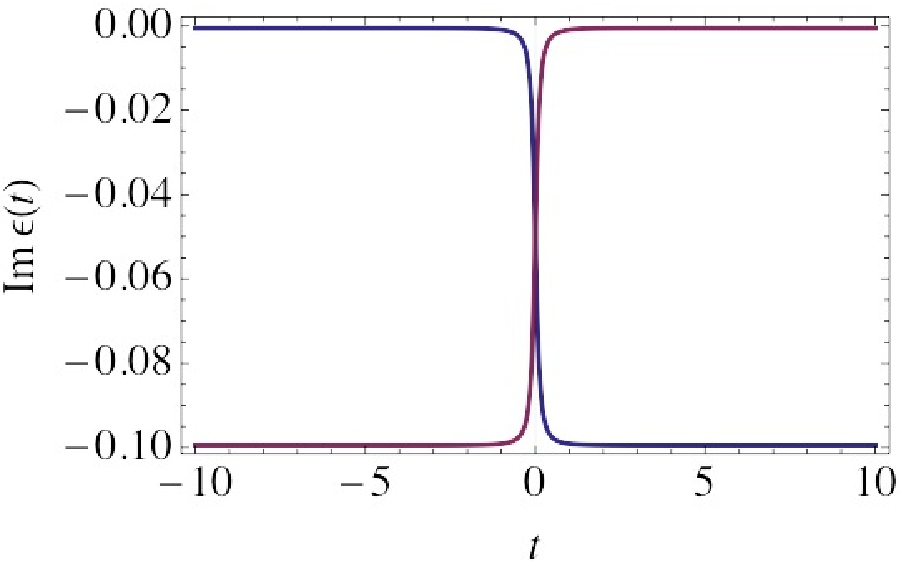}}
\centering\subfigure[]{\includegraphics[width=0.24\textwidth]{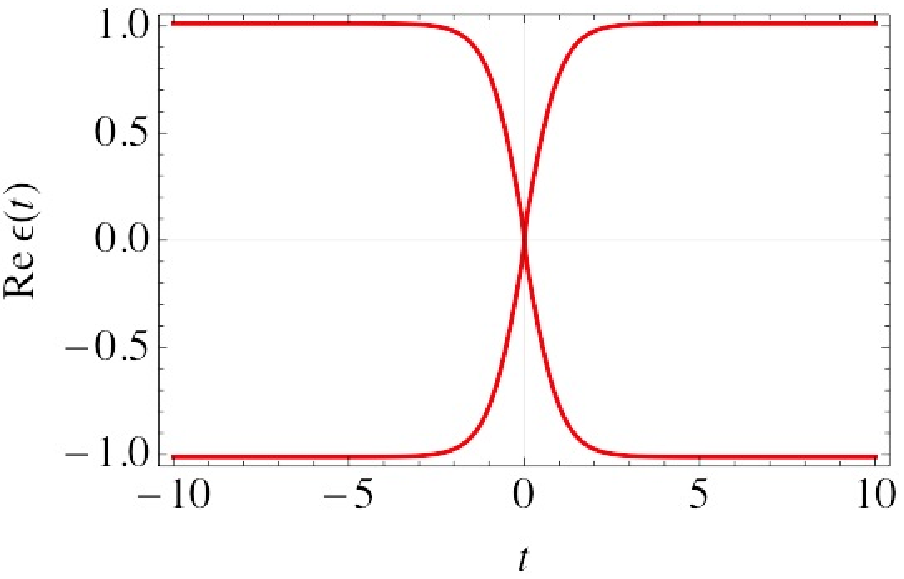}}
\centering\subfigure[]{\includegraphics[width=0.24\textwidth]{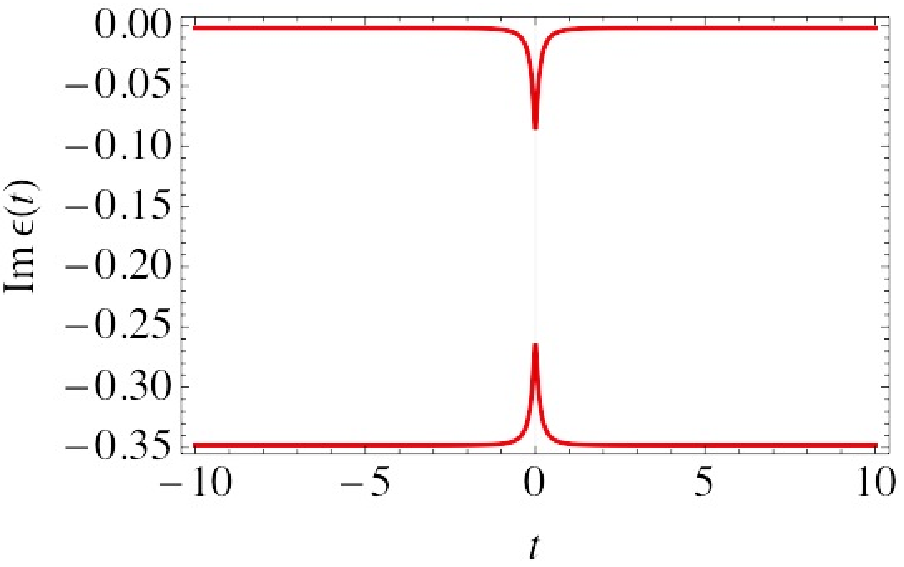}}
\caption{Complex egenenergies  of the Hamiltonian
$H=\binom{\tanh t \ \ \ \ \ \ \ \ \ \ \ \ \ \lambda}{\lambda \ \ \ \ \ \
-(\tanh t+i \beta}$ as a function of time. (a) ${\mathrm{Re}}[\veps(t)]$, and 
(b) ${\mathrm{Im}}[\veps(t)]$, with $\lambda=0.15$ and $\beta=0.2$.
Since $4 \lambda > \beta$, the corresponding level pattern is as
discussed in points (1)-(4). (c) ${\mathrm{Re}}[\veps(t)]$, and (d) 
${\mathrm{Im}}[\veps(t)]$, with $\lambda=0.15$ and $\beta=0.7$.
Since $4 \lambda < \beta$, the corresponding level pattern is as
discussed in point (5) above.}
\label{EV1}
\end{figure}


\subsubsection{St\"uckelberg oscillations with $\lambda$ and the increase
of $P(T;\lambda,\beta)$ with $\beta$}

In the present section we focus on two aspects of $P(T;\lambda,\beta)$ for 
fixed and large $T$.  First we study St\"uckelberg oscillations with 
$\lambda$ for finite decay rate, $\beta > 0$, and then we show that
$P(T;\lambda,\beta)$ can increase with increasing $\beta$.  As will become 
evident, these two aspects of the LZ dynamics are intimately related. 

In Sec.~\ref{SubSec:AS_numerical} we encountered St\"uckelberg
oscillations of $|\psi_1(t)|^2$ with time for $\beta=0$ and fixed
$\lambda$.  Here we show that there are also St\"uckelberg
oscillations of $P(T;\lambda, \beta)$ with varying $\lambda$ and
finite but small $\beta$.  We also show that $P(T;\lambda, \beta)$
increases with $\beta$ for sufficiently large $\beta$.  Using the
analytic expression (\ref{psi1T}) for the wave function $\psi_1(T)$ we
compute $P(T;\lambda,\beta)$, and elucidate its dependence on the
relevant parameters.  The results of this section are illustrated in 
Fig.~\ref{Fig_LZ_TH3} which contains a 3D plot of $P(T;\lambda,\beta)$.
The main features that can be deduced from this figure are outlined in 
the caption. 
In brief, (1) the St\"uckelberg oscillations are quite sizable at 
$\beta=0$ (no decay), and are much milder for small $\beta$, and 
eventually die out at larger $\beta$.  This decay of the oscillations 
can be deduced from the initial Hamiltonian (\ref{THDimensionless});
for large $\beta$, the element $H_{22}(t)$ is dominated by its imaginary 
part. (2) As mentioned in the previous discussion, there are 
cases where  $P(T;\lambda,\beta)$ for fixed $\lambda$ increases 
with $\beta$. This is especially evident when the probability is 
examined for $\lambda$ that corresponds to a minimum of a St\"uckleberg 
oscillation, (e.g $\lambda \approx 0.5$ in Fig.~\ref{Fig_LZ_TH3}).

%

\begin{figure}
\centering
\includegraphics[width=0.5 \textwidth]{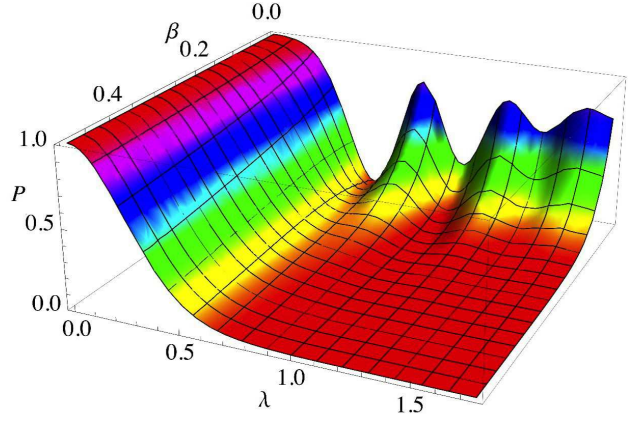}
\caption{(Color online) The probability $P(T; \lambda, \beta)$ based
on Eq.~(\ref{psiTh}) as a function of $\lambda$ and $\beta$ for
$T=10$.  The main features are the St\"uckelberg oscillations at
$\beta=0$, the decrease of the probability with increasing $\beta$ for
small $\lambda$ (e.g., $\lambda = \approx 0.2$), the decay of $P$ with
increasing $\beta$ for moderate $\beta$ (e.g., $\beta = 0.3$), and the
slow {\it increase} of $P$ as as a function of $\beta$ near the minimum of 
the St\"uckelberg oscillations for small $\lambda$ (i.e., 
$\lambda \approx 0.5$).}
\label{Fig_LZ_TH3}
\end{figure}

We can further elaborate on this (somewhat counter-intuitive) result 
with the help of a few two dimensional plots. 
In Fig.~\ref{Fig_LZ_P1lambet}(a) the probability $P(T;\lambda,\beta)$
is plotted as a function of $\lambda$ for $\beta=0$ (dashed curve) and
for $\beta=0.4$ (solid curve) for $T=6$.  For $\beta=0$ the St\"ukelberg
oscillations with $\lambda$ 
are quite violent, but they are also noticeable for finite decay rate 
$\beta=0.4$.  For small $\lambda$, there are cases where the probability 
{\it increases} with $\beta$, as already noted in connection with 
Fig.~\ref{LZ_w_decay_beta_10_l_0.3}.  This remarkable observation is 
further corroborated in Fig.~\ref{Fig_LZ_P1lambet}(b) which displays the
probability $P(T;\lambda,\beta)$ for $T=6$ as function of $\beta$
for fixed $\lambda=0.4$ (dashed curve) and for $\lambda=0.6$ (solid
curve).

\begin{figure}
\centering\subfigure[]{\includegraphics[width=0.45\textwidth]
{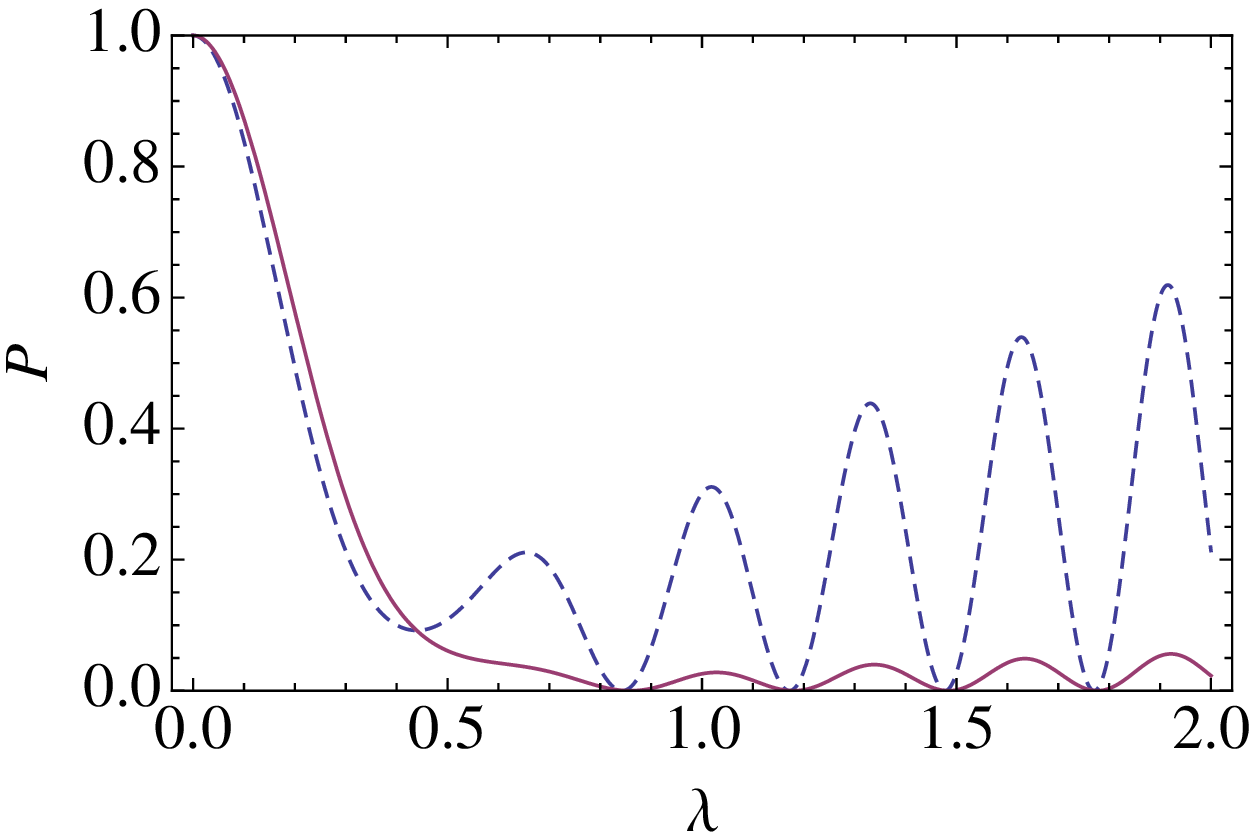}}
\centering\subfigure[]{\includegraphics[width=0.45\textwidth]
{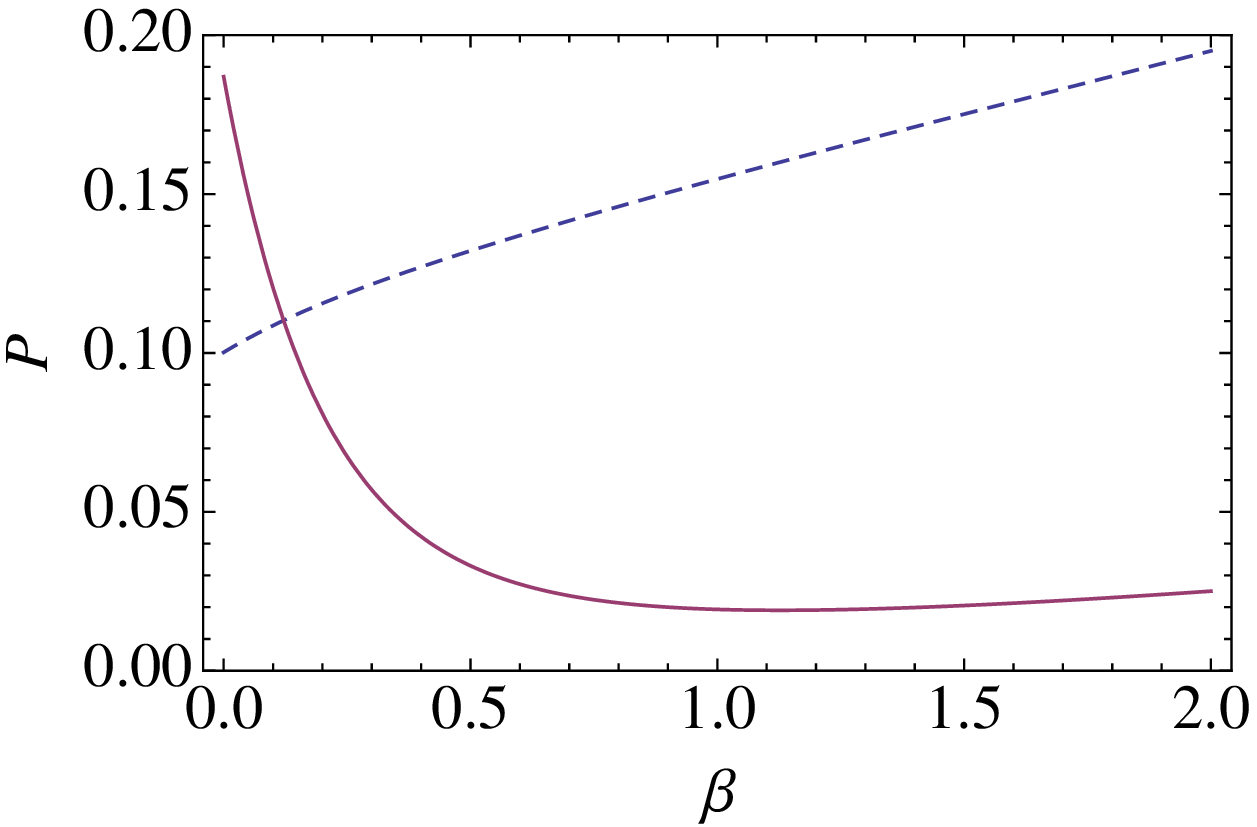}} 
\caption{(a) St\"uckelberg oscillations of $P(T;\lambda, \beta)$ with
$\lambda$.  The probability $P(T;\lambda,\beta)$ for $T=6$ versus
$\lambda$ for $\beta=0$ (dashed curve) and for $\beta=0.4$ (solid
curve).  (b) The probability $P(T;\lambda,\beta)$ versus $\beta$ 
for fixed $\lambda=0.4$ (dashed curve) and for $\lambda=0.6$ (solid 
curve).  For $\lambda=0.4$ (dashed curve), $\beta=0$ is a minimum of 
the first St\"uckelberg oscillation [see part (a)]; hence, 
$P(T;\lambda,\beta)$ increases monotonically with increasing $\beta$.
For $\lambda=0.6$ (solid curve), $\beta=0$ is close to the maximum of 
the first St\"uckelberg oscillation [see part (a)] and 
$P(T;\lambda,\beta)$ decreases at small $\beta$ and then slowly 
increases at higher $\beta$.}
\label{Fig_LZ_P1lambet}
\end{figure}

The probability $P(T; \lambda, \beta) = |\psi_1(T)|^2$, determined
using Eq.~(\ref{psiTh})}, is plotted versus $\lambda$ for $\beta=0$
and $\beta = 0.2$ in Fig.~\ref{Fig_LZ_TH1}(a).  The St\"uckelberg
oscillations for $\beta=0$ (dashed curve) are indeed strong, but for
$\beta = 0.2$ (solid curve) they are subdued, and the probability is
diminished relative to the probability for $\beta=0$.  However,
compared with the first minimum of $P(T;\lambda,\beta)$ for $\beta=0$
at $\lambda \approx 0.55$, the probability {\it increases} with
$\beta$.  We attribute this surprising result to the fact that level
decay induces level crossing.  This is especially pronounced when
$P(T;\lambda,\beta=0)$ is a minimum point in the pattern of
St\"uckleberg oscillations.

\begin{figure}
\centering\subfigure[]{\includegraphics[width=0.45\textwidth]
{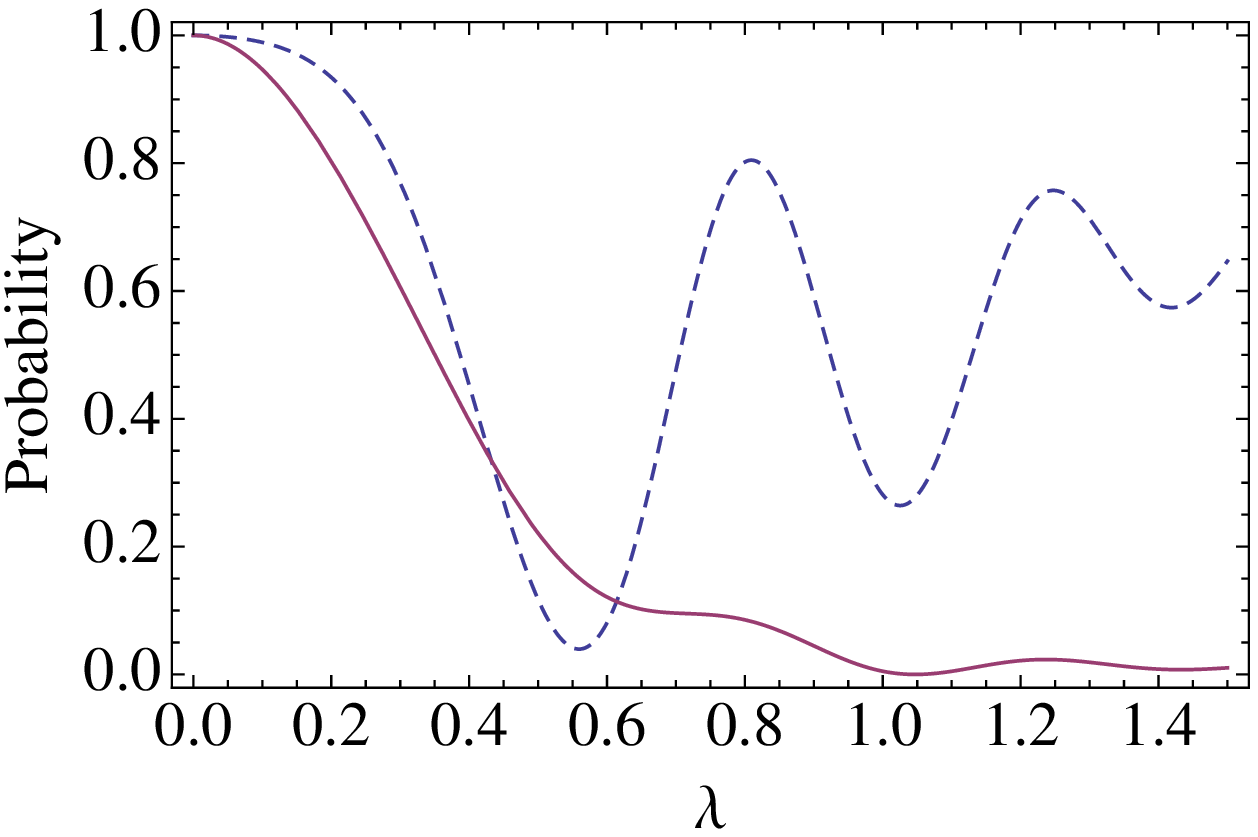}}
\centering\subfigure[]{\includegraphics[width=0.45\textwidth]
{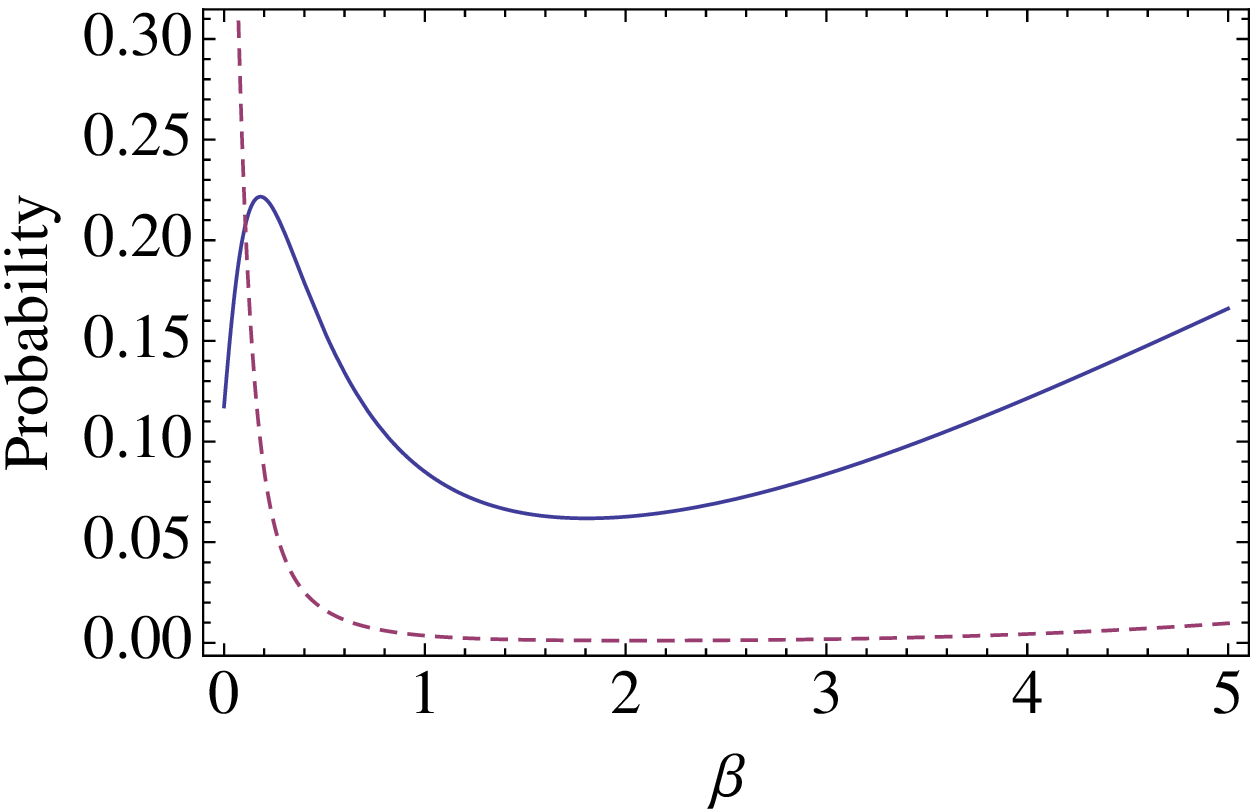}}
\caption{(a) The probability $P(T; \lambda, \beta)$
calculated using Eq.~(\ref{psiTh}) as function of $\lambda$ for
$T=10$, and $\beta=0$ (dashed curve) and $\beta=0.2$ (solid curve).
Note that between $\lambda = 0.45$ and 0.61, the curve with decay is
larger than without decay.
(b) The probability $P(T; \lambda, \beta)$ based on Eq.~(\ref{psiTh})
as function of $\beta$ for $T=10$, and $\lambda=0.5$ (solid curve) and
$\lambda=0.8$ (dashed curve).  For $\lambda = 0.5$ (solid curve),
$\beta=0$ is at a minimum of the first St\"uckelberg oscillation [see
part (a)] and $P(T;\lambda,\beta)$ increases with increasing $\beta$,
then decreases and subsequently slowly increases.  For $\lambda=0.8$
(dashed curve), $\beta=0$ is close to the maximum of the first
St\"uckelberg oscillation [see part (a)] and $P(T;\lambda,\beta)$
decreases with increasing $\beta$ at small $\beta$, and then slowly
increases at higher $\beta$.}
\label{Fig_LZ_TH1}
\end{figure}



Figure~\ref{Fig_LZ_TH1}(b) plots the probability $P(T; \lambda,\beta)$
as function of $\beta$ for $\lambda=0.5$ (solid line) and $\lambda=0.8$ 
(dashed line).  Consider first the solid curve for $\lambda=0.5$, related 
to the discussion of Fig.~\ref{Fig_LZ_TH1}(a).  The curve starts at the
first minimum of the St\"uckelberg oscillations and reaches a local
maximum, after which it decays, as expected for a problem with
increasing decay rate.  However, at higher $\beta$, it approaches and
then crosses the curve $\beta = 4 \lambda$.  As discussed in points
(4) and (5), the probability increases, which is counter-intuitive.
For $\lambda=0.8$ $P(T; \lambda,\beta)$ starts its decay right at the
onset as expected, but again, unexpectedly, it starts to increase at
higher $\beta$ since $\beta$ approaches and crosses the point $\beta =
4 \lambda = 3.2$.  For strong coupling $\lambda$, however, this rise
of $P(T; \lambda, \beta)$ is less visible. 

\section{Numerical Results for the LZ problem with decay of both levels}
\label{SubSec:decay_both_levels}

\begin{figure}
\centering\subfigure[]{\includegraphics[width=0.45\textwidth]
{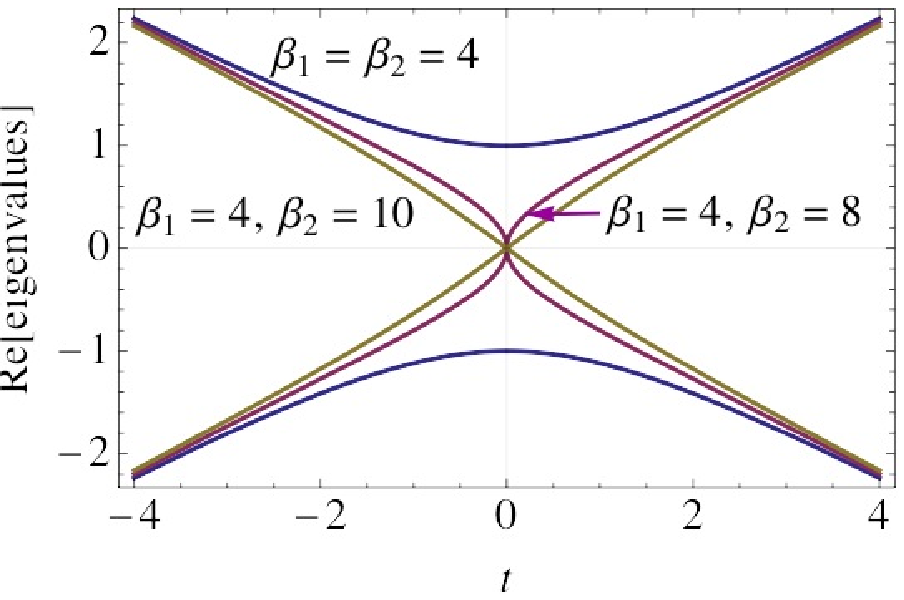}}
\centering\subfigure[]{\includegraphics[width=0.45\textwidth]
{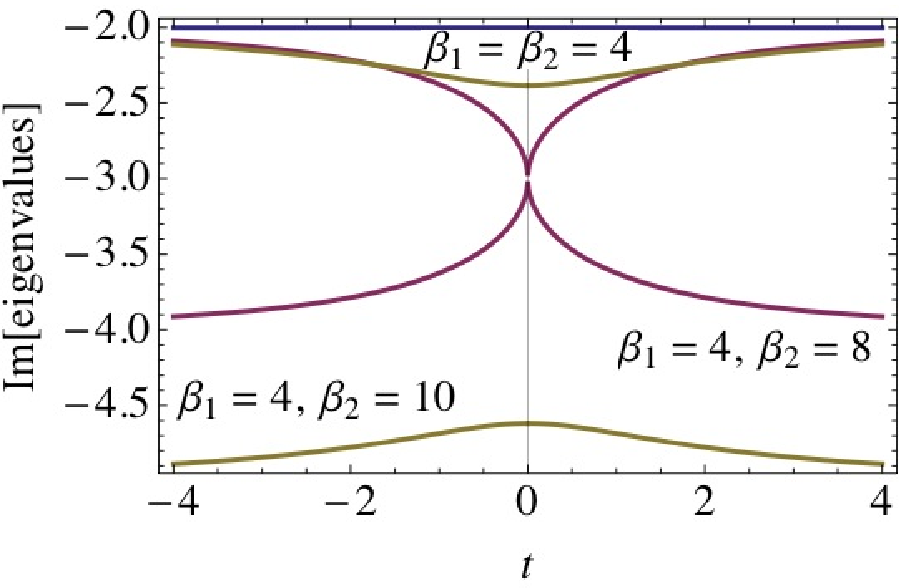}}
\caption{(Color online) For the Hamiltonian in
Eq.~(\ref{Eq:Ht_dim_2decay}) with $\lambda = 1$, (a)
${\mathrm{Re}}[\veps_{1,2}(t)]$ versus time, (b)
${\mathrm{Im}}[\veps_{1,2}(t)]$ versus time.}
\label{Fig_LZ_2decay}
\end{figure}

Systems for which both levels in the LZ dynamics undergo decay to states
outside the two-level manifold exist in nuclear and mesoscopic systems 
\cite{VZ_11}.  From our study in the 
previous sections we learned that the dependence of $P(T;\lambda, 
\beta_1,\beta_2)$ on $\beta_2$ is sometimes not simple.  But its dependence 
on $\beta_1$ is much simpler. Clearly, in the absence of coupling 
($\lambda=0$), the probability $P(T;\lambda, \beta_1,\beta_2)$ decays 
exponentially with $\beta_1$.  As we shall see below, switching on the 
coupling $\lambda$ does not affect
this behavior in any significant way.  We start from the symmetric
form of the Hamiltonian, corresponding to $Z_1(t) = \tfrac{1}{2}(t-i
\beta)$ and $Z_2(t) = -\tfrac{1}{2}(t+i \beta)$ in
Eq.~(\ref{LZgeneral1}).  Such a Hamiltonian corresponds to the case of
spin $S = 1$ states with $M_S = 1$ and $M_S = -1$ in the presence of
an external magnetic field along the $z$-axis whose strength is
changed linearly in time.  Adding decay to both levels yields the
Hamiltonian
\begin{equation}  \label{Eq:Ht_dim_2decay}
    {\cal H}(t) = \left( \! \!\begin{array}{cc}  \tfrac{1}{2}(t-i
    \beta_1) & \lambda \\
    \lambda & -\tfrac{1}{2}(t+i \beta_2) \end{array} \!  \!  \right) ,
\end{equation}
The eigenvalues of this Hamiltonian are
\begin{equation} \label{eigen_sym_2decay}
    \veps_{1,2}(t) = \frac{1}{4} \left[-i (\beta_1 + \beta_2) \pm 
    \sqrt{[2 t -i (\beta_1 - \beta_2)]^2 + 16 \lambda^2}\right] .
\end{equation}
The factor $-i(\beta_1 + \beta_2)$ appears in both eigenvalues and
affects the decay dynamics by introducing exponential decay of the
time-dependent wave function \cite{Burshtein_88}.  When $|\beta_1 -
\beta_2| \ge 4\lambda$ the eigenvalues cross.
Figure~\ref{Fig_LZ_2decay} plots the real and imaginary parts of the
eigenvalues for $\lambda = 1$ for three different cases of decay
constant pairs: (1) $\beta_1 = \beta_2 = 4$, where there is an avoided
crossing, (2) the borderline case, $2\beta_1 = \beta_2 = 8$ so that
$|\beta_1 - \beta_2| = 4 \lambda$, which is the onset of crossing, and
(3) $\beta_1 = 4$ and $\beta_2 = 10$.  In the examples of the dynamics
that follow, the consequences of the avoided crossing or crossing will
be very noticeable.

\begin{figure}
\centering
\includegraphics[width=0.45 \textwidth]{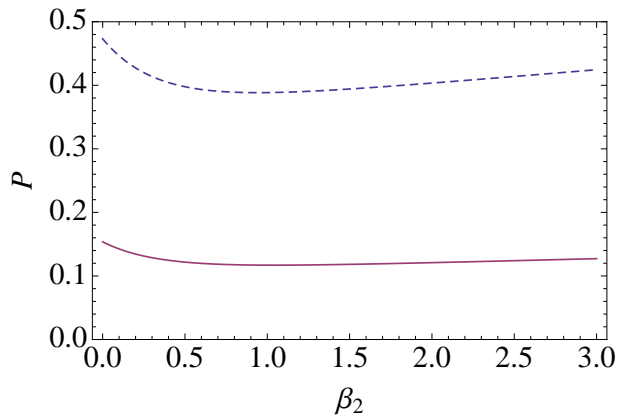}
\caption{(Color online) The probability $P(T;\lambda,
\beta_1,\beta_2)$ for $T=6$ as a function of $\beta_2$ for fixed
$\lambda=0.4$ and for two values of $\beta_1$, $\beta_1=0$ (dashed
curve) and $\beta_1=0.1$ (solid curve).  The effect of level crossing
reflected by the slow increase of $P(T;\lambda, \beta_1,\beta_2)$ with
$\beta_2$ is clearly seen for $\beta_1 = 0$ while it is hardly visible
even at small $\beta_1$, e.g., $\beta_1 = 0.1$.}
\label{Fig_Pbeta1_beta2}
\end{figure}

When $\beta_1 \ne 0$, it is expected that
$P(T;\lambda,\beta_1,\beta_2)$ decay exponentially with $\beta_1$.
More precisely, assume for the moment that there is no level
interaction, e.g., if $\lambda=0$, then, $\psi_1(t) = e^{i
(t^2-T^2)/4} e^{-\beta_1(t+T)/2}$.  Thus, the survival probability is
$P(T;\lambda,\beta_1,\beta_2) =|\psi_1(T)|^2 = e^{- \beta_1T}$.
Switching on the coupling, $\lambda \ne 0$, affects the above result
(valid for $\lambda=0$) due to depopulation of level 1. In particular, 
it leads to the possible increase of  $P(T;\lambda,\beta_1,\beta_2)$ with 
$\beta_2$, but, as we shall see in Fig.~\ref{Fig_Pbeta1_beta2}, in a
much less significant fashion than in the former case where $\beta_1=0$.  
Thus, when $\beta_1, \beta_2 \ne 0$ the exponential decay is the dominant 
feature of the dynamics.
 
\begin{figure}
\centering\subfigure[]{\includegraphics[width=0.45\textwidth]
{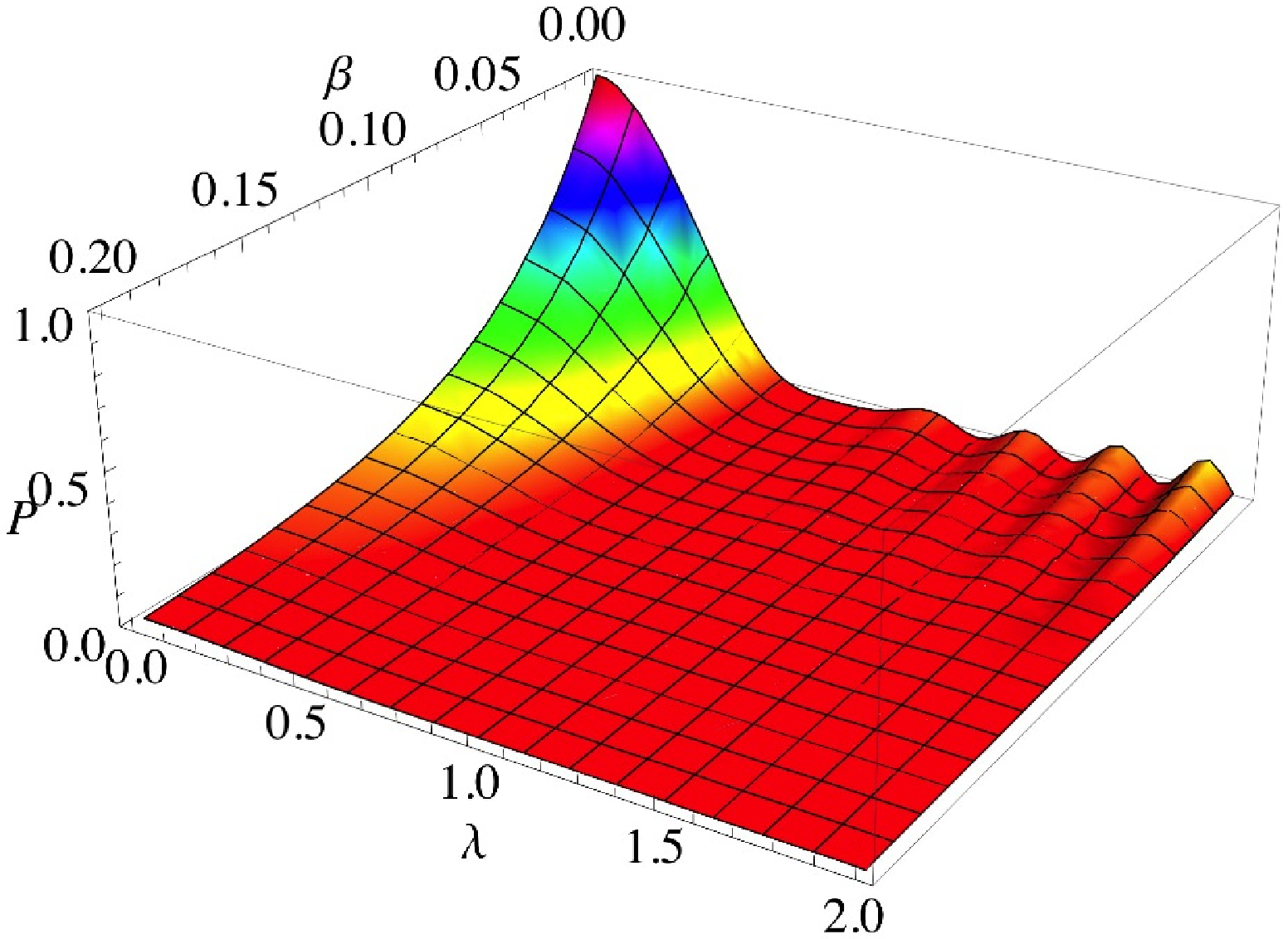}}
\centering\subfigure[]{\includegraphics[width=0.45\textwidth]
{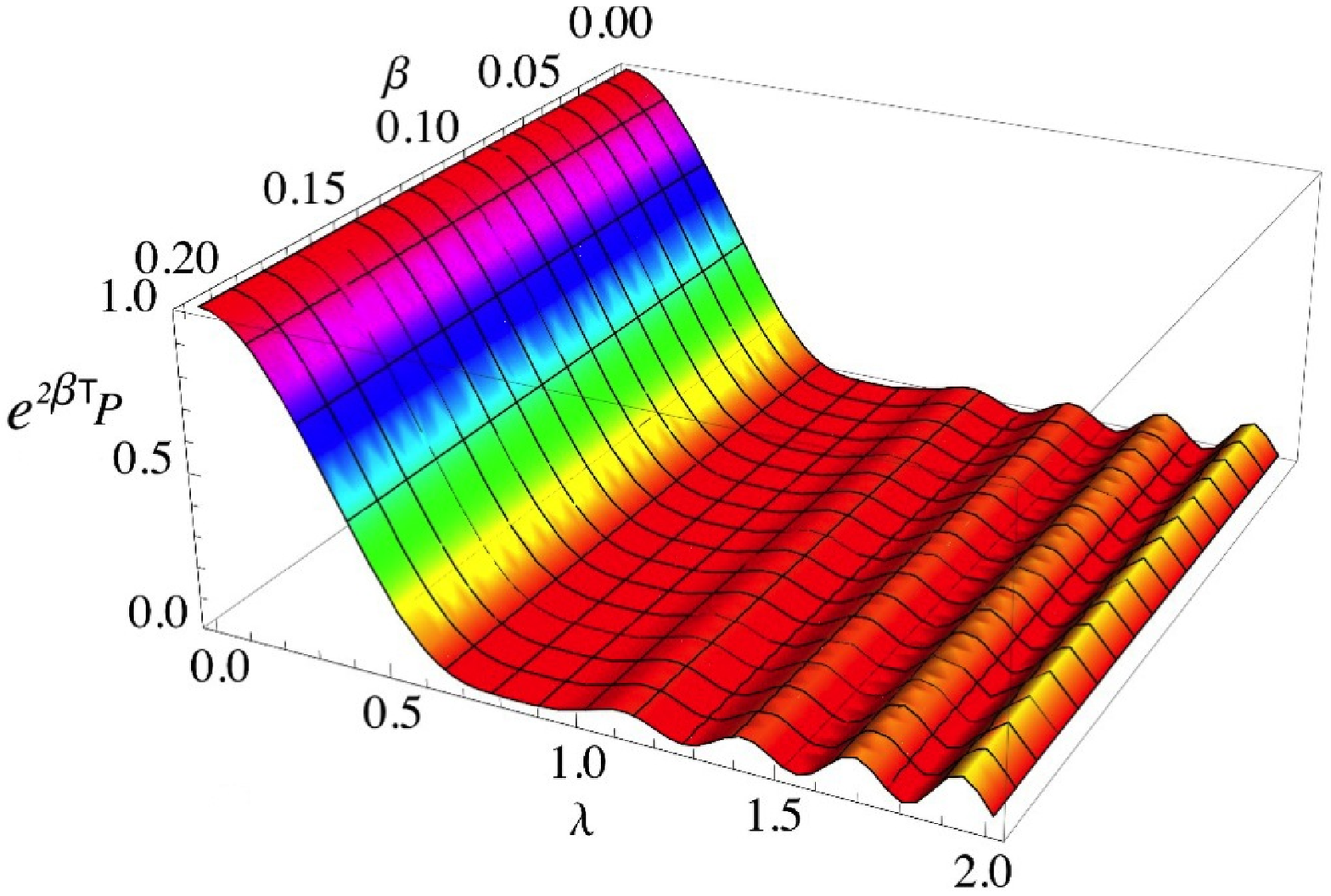}} 
\caption{(Color online) (a) The probability $P(T; \lambda,
\beta_1,\beta_2)$, calculated using the Hamiltonian
Eq.~(\ref{Eq:Ht_dim_2decay}), as a function of $\lambda$ and $\beta
\equiv \beta_1 = \beta_2$ for $T=10$.  (b) $e^{\beta T} \, P(T;
\lambda, \beta, \beta)$ calculated using the Hamiltonian
Eq.~(\ref{Eq:Ht_dim_2decay}) as a function of $\lambda$ and $\beta
\equiv \beta_1 = \beta_2$ for $T=10$.}
\label{Fig_LZ_2decay_lambda_beta}
\end{figure}

Figure~\ref{Fig_LZ_2decay_lambda_beta}(a) shows the probability $P(T;
\lambda, \beta_1,\beta2)$ for the case where $\beta_1 = \beta_2 \equiv
\beta$, as a function of $\lambda$ and $\beta$.  The probability
decays exponentially in both $\lambda$ and $\beta$, but the physics of
each decay in each variable is distinct.  Decay with $\lambda$,
accompanied by St\"uckelberg oscillations for small $\beta$ which
reflects the interference effect in the avoided crossing dynamics, is
due to the avoided crossing, whereas the decay with $\beta$ reflects
the exponential factor $e^{- \beta T}$ as discussed above.  To show
this, we multiply the probability by $e^{\beta T}$ and plot $e^{ \beta
T} \, P(T; \lambda, \beta, \beta)$ as a function of $\lambda$ and
$\beta \equiv \beta_1 = \beta_2$ in
Fig.~\ref{Fig_LZ_2decay_lambda_beta}(b).  This kind of plot was first
suggested in Ref.~\cite{Burshtein_88}.  The figure clearly shows that
$e^{ \beta T} \, P(T; \lambda, \beta, \beta)$ is independent of
$\beta$.  Of course, for the case $\beta_1=\beta_2=\beta$, this result
is expected because then the decay term enters as $-i (\beta/2) {\bf
I}_{2 \times 2}$, therefore $\psi_1(T,\beta)=\exp^{-(\beta/2)T}
\psi_1(T,0)$.

Finally we address the question of how $\beta_1 \ne 0$ affects the 
observation that $P(T; \lambda, \beta_1=0, \beta_2)$ might increase with 
$\beta_2$. Figure \ref{Fig_Pbeta1_beta2} shows the probability 
$P(T;\lambda,\beta_1,\beta_2)$ as function of $\beta_2$ for fixed 
$\lambda=0.4$ for two values of $\beta_1$, $\beta_1=0$ (dashed curve) and 
$\beta_1=0.1$ (solid curve).  The effect of level crossing is reflected 
by the slow increase of $P(T;\lambda, \beta_1,\beta_2)$ with $\beta_2$ is 
clearly seen for $\beta_1 = 0$ while it is hardly visible for $\beta_1 
= 1$.

\section{Landau--Zener Problem with dephasing}
\label{Sec:dephasing}

Dephasing is one of the causes of decoherence of a quantum system, and
is due to the interaction of the system with its environment (see
Sec.~\ref{Sec:Intro}).  Dephasing results in the scrambling of the
phases of the amplitudes appearing in the system wave function.  In
the context of magnetic resonance phenomena, decay and dephasing are
often called $T_1$ and $T_2$ processes respectively.  In this section
we describe an approach for treating the LZ problem with dephasing
which uses a stochastic Schr\"{o}dinger--Langevin differential
equation approach.  We also relate this to the master equation
(density matrix) approach, at least for Gaussian white noise (we also
consider Gaussian colored noise).  For Gaussian white noise, the 
stochastic Schr\"{o}dinger--Langevin  approach is equivalent to a 
master equation approach with Lindblad terms \cite{vanKampenBook}.  
We shall calculate the average over stochastic realizations of the 
LZ survival probability, $\overline{P(t)}$, the standard deviation 
of the probability, $\Delta P(t) = \sqrt{\overline{P(t)^2} - 
(\overline{P(t)})^2}$, and the distribution ${\cal D}[P(T)]$ of the 
probability $P(T)$ at the final time, and analyze the dependence on 
the LZ parameters and the dephasing strength.

\subsection{Analogy with spin 1/2 particle in a stochastic magnetic field}

It is useful to use the analogy of a spin 1/2 particle under the
influence of a time-dependent stochastic magnetic field to exemplify
the role of dephasing in the LZ problem.  Following
Refs.~\cite{STB_2013, Rammer, Efrat2}, the bare LZ Hamiltonian ${\cal
H}_0$ is a $2$$\times$$2$ matrix that is formally written as ${\cal
H}_0 = {\bm \sigma} \cdot {\bf B}_0(t)$, where ${\bf B}_0$ is the
intrinsic ``magnetic field".  Interaction with the environment is
modeled using a Hamiltonian ${\cal H}_1 = {\bm \sigma} \cdot {\bf
b}(t)$ where ${\bf b}(t)$ is the external stochastic ``magnetic
field".  For $T_2$ dephasing processes, we take ${\bf b}(t) = \xi(t)
\hat {\bf z}$ where $\xi(t)$ is white noise.  The average over the
noise fluctuations and the second moment are given by
\begin{equation}  \label{white_noise}
    \overline{\xi(t)} = 0, \quad \overline{\xi(t) \xi(t')} =
    \xi_0^2 \, \delta(t-t') ,
\end{equation}
where $\xi_0$ is the volatility (the stochastic field strength) which
is inversely proportional to the dephasing time $\tau_\phi$,
$\overline{(\ldots)}$ denotes the stochastic average, and
$\delta(\bullet)$ is the Dirac $\delta$ function.  The white noise,
$\xi(t)$, can be written as the time derivative of the Wiener process,
$\xi(t) = d w(t)/dt$, or more formally, the Wiener process $w(t)$ is
the integral of the white noise.

As before, the initial state of the spin at $t=-T$ is
$\psi_1(-T)=|\! \uparrow \rangle=\binom{1}{0}$, and we seek the
probability $P(T)$ that it will stay at a state $|\! \uparrow \rangle$ at
$t = T$.  Our approach is to numerically solve the time-dependent
Schr\"odinger with a stochastic term proportional to $w(t)$.  Since
$w(t)$ is a stochastic process, $P(t)$ is also, and it has a
distribution ${\cal D}[P(t)]$.  Usually, interest is focused on the
averaged probability, $\overline{P}=\int_0^1 {\cal D}(P) dP$ at any
given time, and in particular, the final time.  More information,
however, is encoded in the distribution ${\cal D}(P)$ of the
probability, and this is less well-studied.

\subsection{Stochastic time-dependent Schr\"odingier equation}

There are several ways of modeling stochastic processes,
including a master equation method \cite{master_eq}, a Monte Carlo
wave-function method \cite{Molmer_93}, or a stochastic differential
equations method.  Here, we model dephasing using stochastic
differential equations \cite{vanKampenBook, Kloeden, Kloeden_03,
Gardiner}.  We briefly elaborate on the time-dependent Schr\"{o}dingier 
equation for the LZ problem with a stochastic term that models 
dephasing processes, and its solution.  For the Hamiltonian in 
Eq.~(\ref{Eq:Ht'_dim}), the stochastic equations can be written as
\begin{subequations}  \label{SL_stoch}
\begin{equation}  \label{SL1_stoch}
    {\dot \psi}_{1}(t) = -i \lambda \psi_{2}(t) + \xi_0 \xi(t) 
    \psi_1(t) - \frac{\xi_0^2}{2} \psi_1(t) ,
\end{equation}
\begin{equation}  \label{SL2_stoch}
    {\dot \psi}_{2}(t) = -i [\lambda \psi_{1}(t) + z(t) \psi_{2}(t)]
    - \xi_0 \xi(t) \psi_2(t) - \frac{\xi_0^2}{2} \psi_2(t) ,
\end{equation}
\end{subequations}
where $z(t) = -t/2$ and $\xi_0$ is a dimensionless volatility which is
inversely proportional to the dimensionless dephasing time
$\tau_\phi$.  These equations can be rewritten in the notation of
stochastic differential equations \cite{Kloeden, Kloeden_03, Gardiner}
as
\begin{subequations}  \label{SL_stoch'}
\begin{equation}  \label{SL1_stoch'}
    d \psi_{1}(t) = -i \lambda \psi_{2}(t) dt - \frac{\xi_0^2}{2}
    \psi_1(t) dt + \psi_1(t) \, dw ,
\end{equation}
\begin{equation}  \label{SL2_stoch'}
    d \psi_{2}(t) = -i [\lambda \psi_{1}(t)dt  + z(t) 
    \psi_{2}(t)  dt] - \frac{\xi_0^2}{2} \psi_2(t) dt
    -  \psi_2(t) \, dw ,
\end{equation}
\end{subequations}
where $w(t)$ is the Wiener process, i.e., $\xi(t) = d w(t)/dt$.  The
$\xi_0^2$ terms in these equations insure unitarity
\cite{vanKampenBook}.  For any fixed realization of the stochastic
process, the equations are solved to yield the two component spinor
$\binom {\psi_1(t)}{\psi_2(t)}$ and the survival probability at time
$t$ is $P(t)=|\psi_1(t)|^2$.  The distinction as compared with the
deterministic case $\xi_0=0$ is that now $P(t)$ is a random function
with distribution ${\cal D}[P(t)]$ (see
Sec.~\ref{SubSec:LZ_dephasing_numerical}).

Equations (\ref{SL_stoch}) [or (\ref{SL_stoch'})] are a special case
of the Schr\"{o}dinger--Langevin equation \cite{vanKampenBook},
\begin{equation}  \label{Schr_Langevin}
    {\dot \psi} = -i {\cal H} \psi + \xi_0 \xi(t) {\cal V} \psi -
    \frac{\xi_0^2}{2} {\cal V}^{\dag} {\cal V} \psi .
\end{equation}
In our case, ${\cal V} = \sigma_z$, and $\psi$ is a two component
spinor.  Equation (\ref{Schr_Langevin}) can be generalized to include
{\em sets} of operators ${\cal V}_j$, stochastic processes $w_j(t)$,
and volatilities $w_{0,j}$, to obtain the {\em general 
Schr\"{o}dinger--Langevin equation},
\begin{equation}  \label{gen_Schr_Langevin}
    {\dot \psi} = -i {\cal H} \psi + \sum_j \left(\xi_{0,j} \xi_j(t)
    {\cal V}_j \psi - \frac{\xi_{0,j}^2}{2} {\cal V}^{\dag}_j {\cal
    V}_j \psi \right) .
\end{equation}
The average over stochasticity obtained using
Eq.~(\ref{gen_Schr_Langevin}) will be equal the result obtained using
a Markovian quantum master equation for the density matrix $\rho(t)$
with Lindblad operators ${\cal V}_j$ \cite{vanKampenBook, master_eq},
\begin{equation}  \label{Eq:master}
    {\dot \rho} = -i [{\cal H}, \rho(t)] + \frac{1}{2} \sum_j w_{0,j}^2
    \left(2{\cal V}_j \rho(t) {\cal V}^{\dag}_j - \rho(t) {\cal
    V}^{\dag}_j {\cal V}_j - {\cal V}^{\dag}_j {\cal V}_j \rho(t)
    \right) .
\end{equation}
A numerical demonstration of the equivalence is presented in
Ref.~\cite{Band_RWA}.  However, the master equation will not yield the
variance or the statistics or the distribution, quantities that can be
obtained from the stochastic Schr\"{o}dinger--Langevin equation
approach.

There are many other kinds of stochastic processes.  For example, a
well-known stochastic process is Brownian motion, also known as
Gaussian colored noise and the Ornstein--Uhlenbeck process
\cite{OU_30}.  For this type of stochastic dephasing process process,
the stochastic differential equations are,
\begin{subequations}  \label{SL_stoch_OU}
\begin{equation}  \label{SL1_stoch_OU}
    d \psi_{1}(t) = -i \lambda \psi_{2}(t) dt + \psi_1(t) \, {\cal O}(t) ,
\end{equation}
\begin{equation}  \label{SL2_stoch_OU}
    d \psi_{2}(t) = -i [ \lambda \psi_{1}(t)dt  + z(t) 
    \psi_{2}(t)  dt] -  \psi_2(t) \, {\cal O}(t) ,
\end{equation}
\begin{equation}  \label{SL3_stoch_OU}
    d {\cal O}(t) = \vartheta [\mu - {\cal O}(t)]
    \psi_{2}(t)  dt + \sigma dW(t) ,
\end{equation}
\end{subequations}
where the mean autocorrelation function of the Ornstein--Uhlenbeck 
process are
\begin{equation}  \label{OU_noise}
    \overline{{\cal O}(t)} = {\cal O}_0 \, e^{-\vartheta t} + \mu (1 -
    e^{-\vartheta t}) , \quad \overline{{\cal O}(t) \, {\cal O}(t')} =
    \frac{\sigma^2}{2\vartheta} e^{-\vartheta (t+t')} [e^{\vartheta \,
    {\mathrm{min}}(t,t')} - 1] ,
\end{equation}
$\vartheta$ is the mean reversion rate of the Ornstein--Uhlenbeck
process ${\cal O}(t)$, $\sigma$ is the volatility, and $\mu$ is the mean,
which we take to vanish, $\mu = 0$; we also take ${\cal O}_0 = 0$.  This
process yields a non-Markovian master equation for the density matrix
of the system.

\subsection{The Landau--Zener problem with dephasing: Numerical results}
\label{SubSec:LZ_dephasing_numerical}

\begin{figure}
\centering\subfigure[]{\includegraphics[width=0.45\textwidth]
{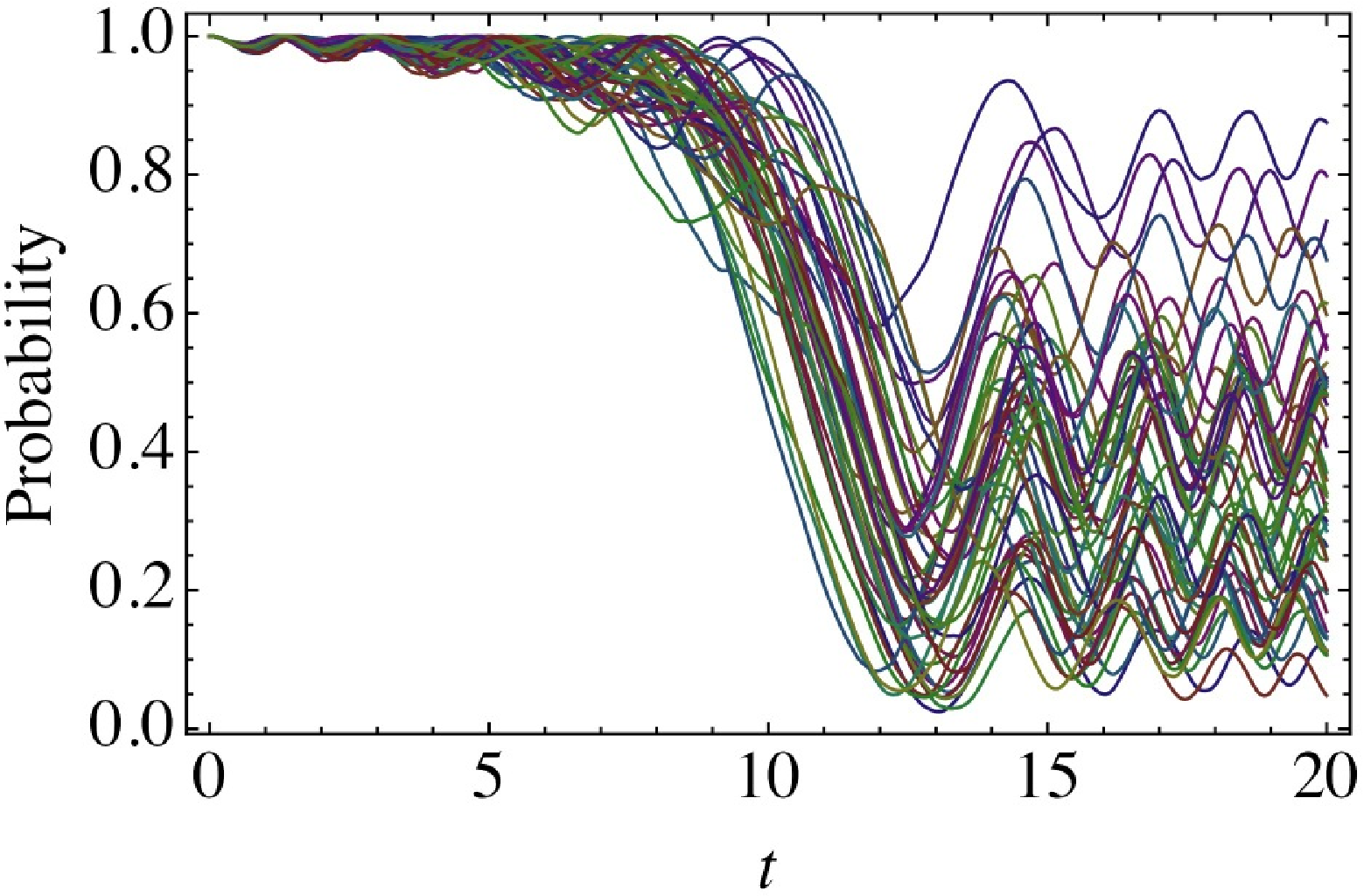}}
\centering\subfigure[]{\includegraphics[width=0.45\textwidth]
{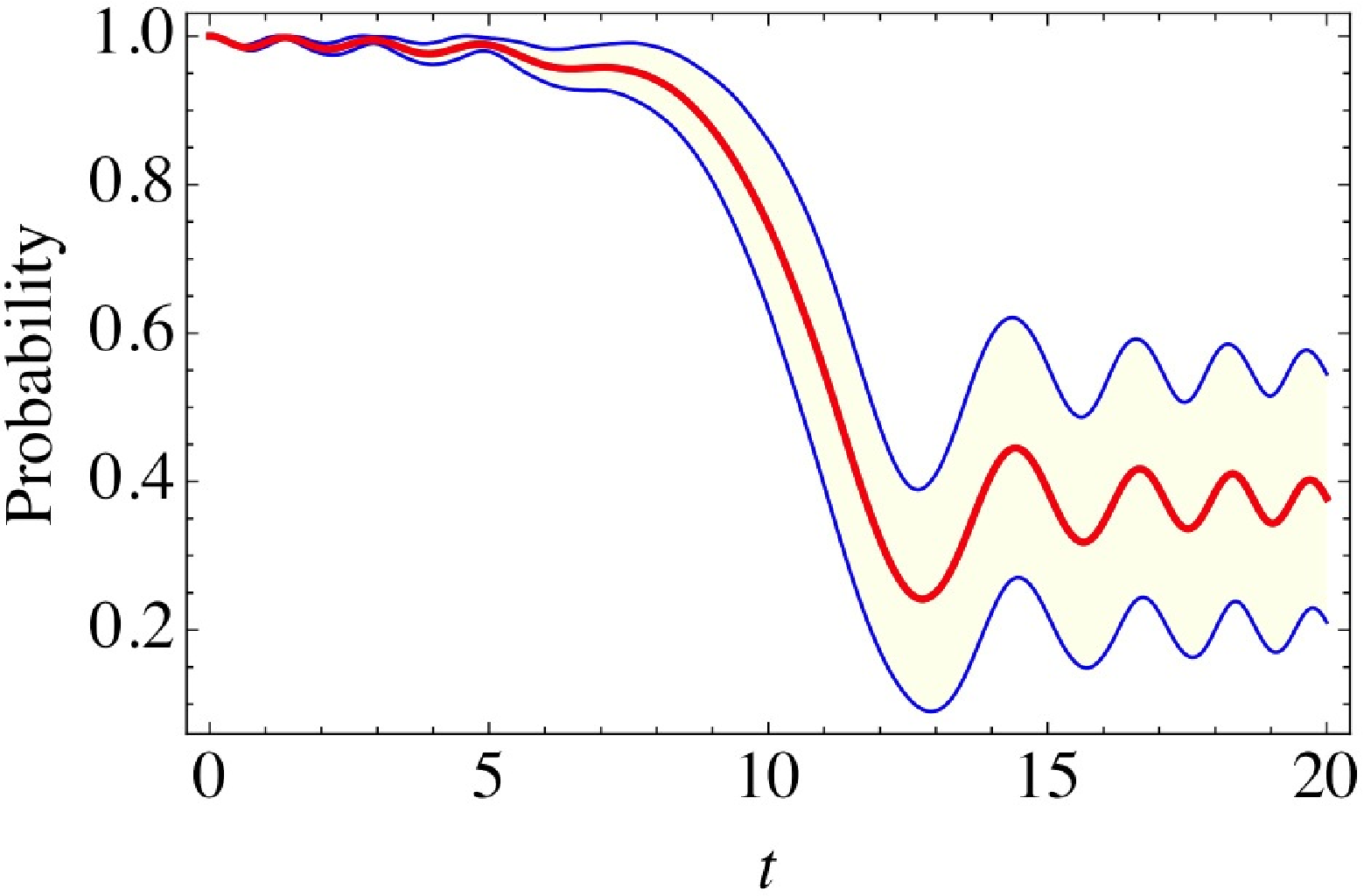}}
\caption{(Color online) Landau--Zener problem with dephasing.  (a) The
probability $P(t;\lambda) = \psi_1^*(t) \psi_1(t)$ versus time for
$\lambda=0.3$ and $T=10$ (i.e., the total elapsed time is $2T = 20$)
for fifty paths (stochastic realizations) with dimensionless
volatility $\xi_0 = 0.2$ [see Eq.~(\ref{white_noise})].  (b) Average
$\overline{P_1(t;\lambda)} = \overline{\psi_1^*(t) \psi_1(t)}$ with
300 stochastic realizations and the average plus and minus standard
deviation of the probability versus time.}
\label{Fig_LZ_stochastic_L_z_0.2}
\end{figure}

Figure~\ref{Fig_LZ_stochastic_L_z_0.2} shows the results of
calculations implemented with the Mathematica 9.0 built-in command
{\em ItoProcess} \cite{ItoProcess} carried out using
Eqs.~(\ref{SL_stoch'}).  We took $\lambda=0.3$ and $T = 10$, so we can
directly compare with the deterministic results (without noise, i.e.,
with $\xi_0 = 0$) shown as the dashed curves in
Fig.~\ref{LZ_w_decay_beta_10_l_0.3}(b).  We had to shift the time so
that we start the process at $t = 0$, rather than $t = -T$, and end it
at $t = 2T$, in order to get ItoProcess to work.
Figure~\ref{Fig_LZ_stochastic_L_z_0.2}(a) plots fifty stochastic
realizations of the survival probability $P(t;\lambda) = \psi_1^*(t)
\psi_1(t)$ versus time for relatively weak disorder ($\xi_0 = 0.2$)
and Fig.~\ref{Fig_LZ_stochastic_L_z_0.2}(b) plots the averaged
probability $\overline{P(t)} = \overline{\psi_1^*(t) \psi_1(t)}$ (red
curve), and mean values plus and minus the standard deviations (blue
curves) versus time for 300 realizations.  Unitarity, i.e., $\langle
\psi(t) |\psi(t) \rangle = 1$ is preserved for each path
(realization), as insured by the $\xi_0^2$ terms in
Eqs.~(\ref{SL_stoch'}).  Clearly, the mean $\overline{P(t)}$ is very
close to the probability without noise shown in
Fig.~\ref{LZ_w_decay_beta_10_l_0.3}(b) [whose analytic form is given
in terms of Eqs.~(\ref{BasicAS1}) and (\ref{BasicAS2})], despite the
fact that the standard deviation at large times is as large as the
mean (the dephasing is not small in this sense).  The evolution of the
standard deviation grows with time but saturates at large times.

\begin{figure}
\centering
\includegraphics[width=0.5\textwidth]{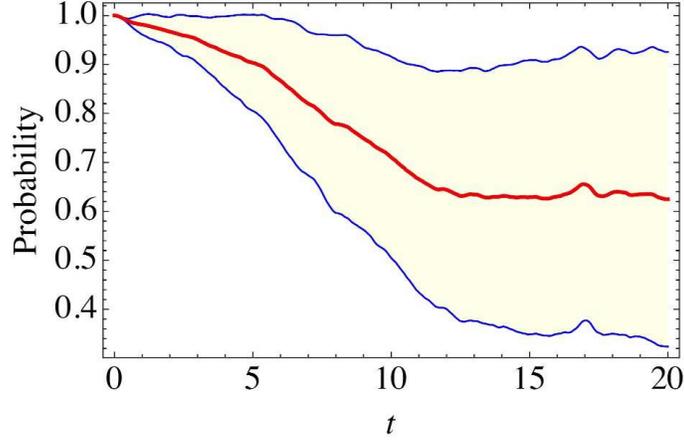}
\caption{(Color online) Same as Fig.~\ref{Fig_LZ_stochastic_L_z_0.2},
except with $\xi_0 = 1.0$.  Average $\overline{P_1(t;\lambda)}$ and 
standard deviation over the 300 realizations.  Compared with
Fig.~\ref{LZ_w_decay_beta_10_l_0.3}(b) (dashed blue curve), where
$P(T=10,\xi_0=0)=0.3$, the noise-averaged probability for high
dephasing rate is much higher, $\overline{P}(t_f;\xi_0=1)=0.6$ 
where $t_f = 2T = 20$.}
\label{Fig_LZ_stochastic_L_z_s_1}
\end{figure}

Let us now consider the stochastic dynamics in the strong
system-environment coupling regime.
Figure~\ref{Fig_LZ_stochastic_L_z_s_1} shows the results for $\xi_0 =
1.0$ and $\lambda = 0.3$.  The mean probability, $\overline{P(t)}$, is
significantly higher and very different in shape than the probability
shown in Fig.~\ref{LZ_w_decay_beta_10_l_0.3}(b).  This is a general
trend of strong dephasing for arbitrary $\lambda$ (see below).  At the
final time we find, $\overline{P(\xi_0=1.0)}-P(\xi_0=0) \approx
0.62-0.35$ with a standard deviation of about 0.3.  Furthermore, for
strong system-environment coupling, the dephasing almost completely
attenuates the interference, which is so significant for the transition
with $\lambda = 0.3$ and no dephasing.

\begin{figure} [!ht]
\centering\subfigure[]{\includegraphics[width=0.45\textwidth]
{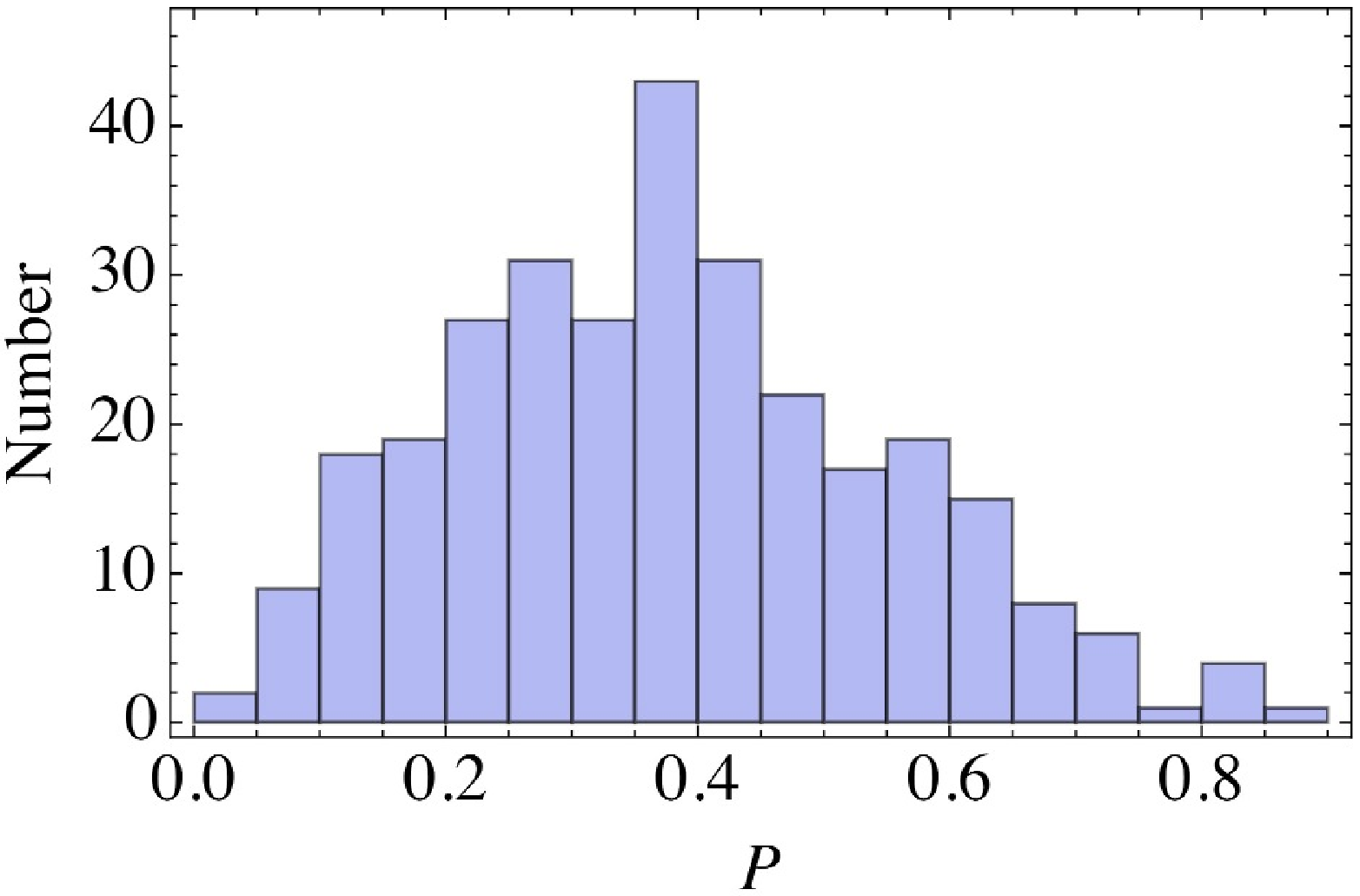}}
\centering\subfigure[]{\includegraphics[width=0.45\textwidth]
{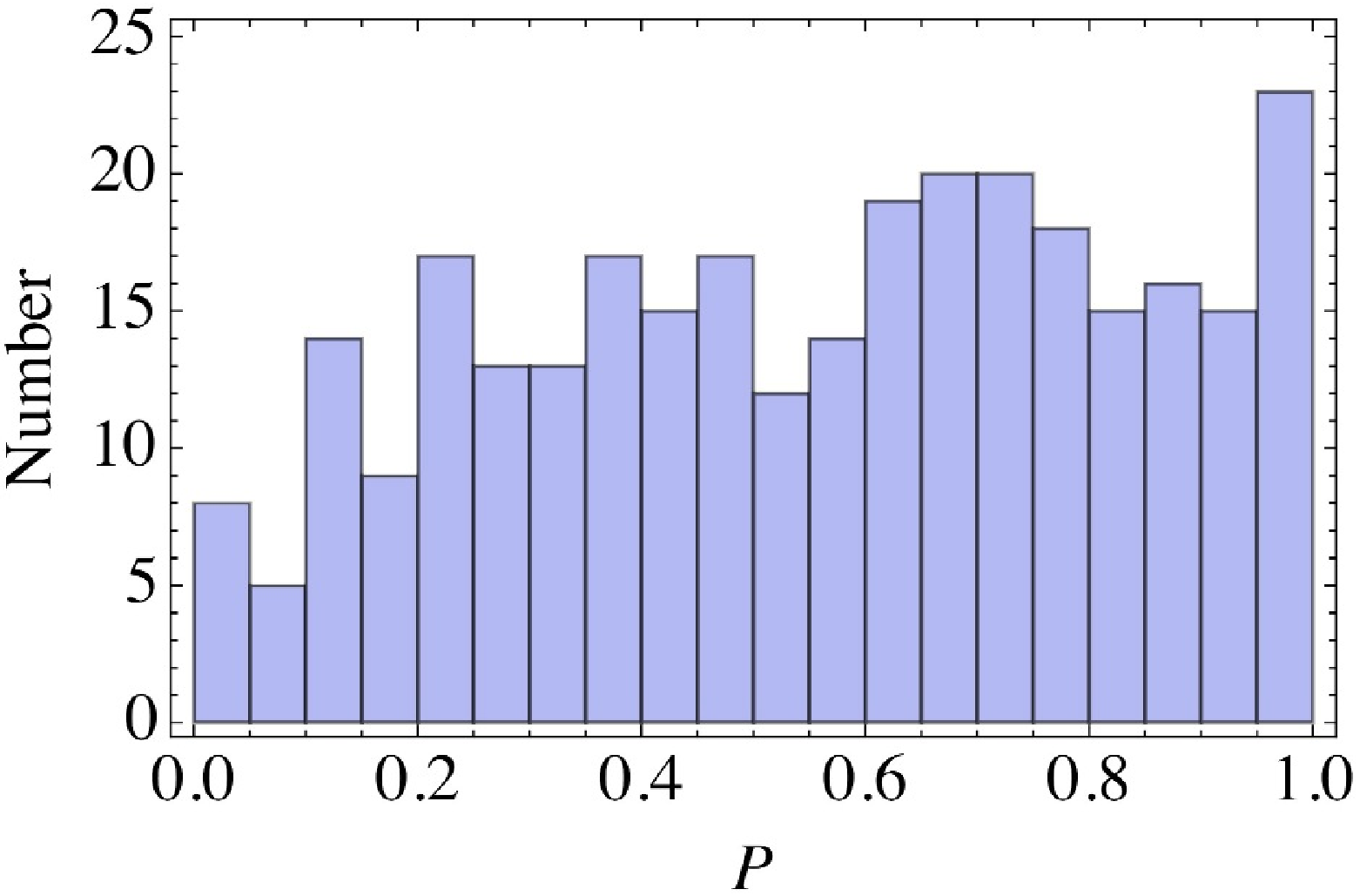}} \caption{(Color online) Histograms of $P(T;\lambda) =
\psi_1^*(T) \psi_1(T)$ with dephasing.  We used $\lambda = 0.3$,
$T=10$ (the total elapsed time is $2T = 20$) with 300 paths
(stochastic realizations).  (a) $\xi_0 = 0.2$ and (b) $\xi_0 = 1$.}
\label{Fig_LZ_stochastic_L_z_Histogram}
\end{figure}

\begin{figure} [!hb]
\centering\subfigure[]{\includegraphics[width=0.32\textwidth]
{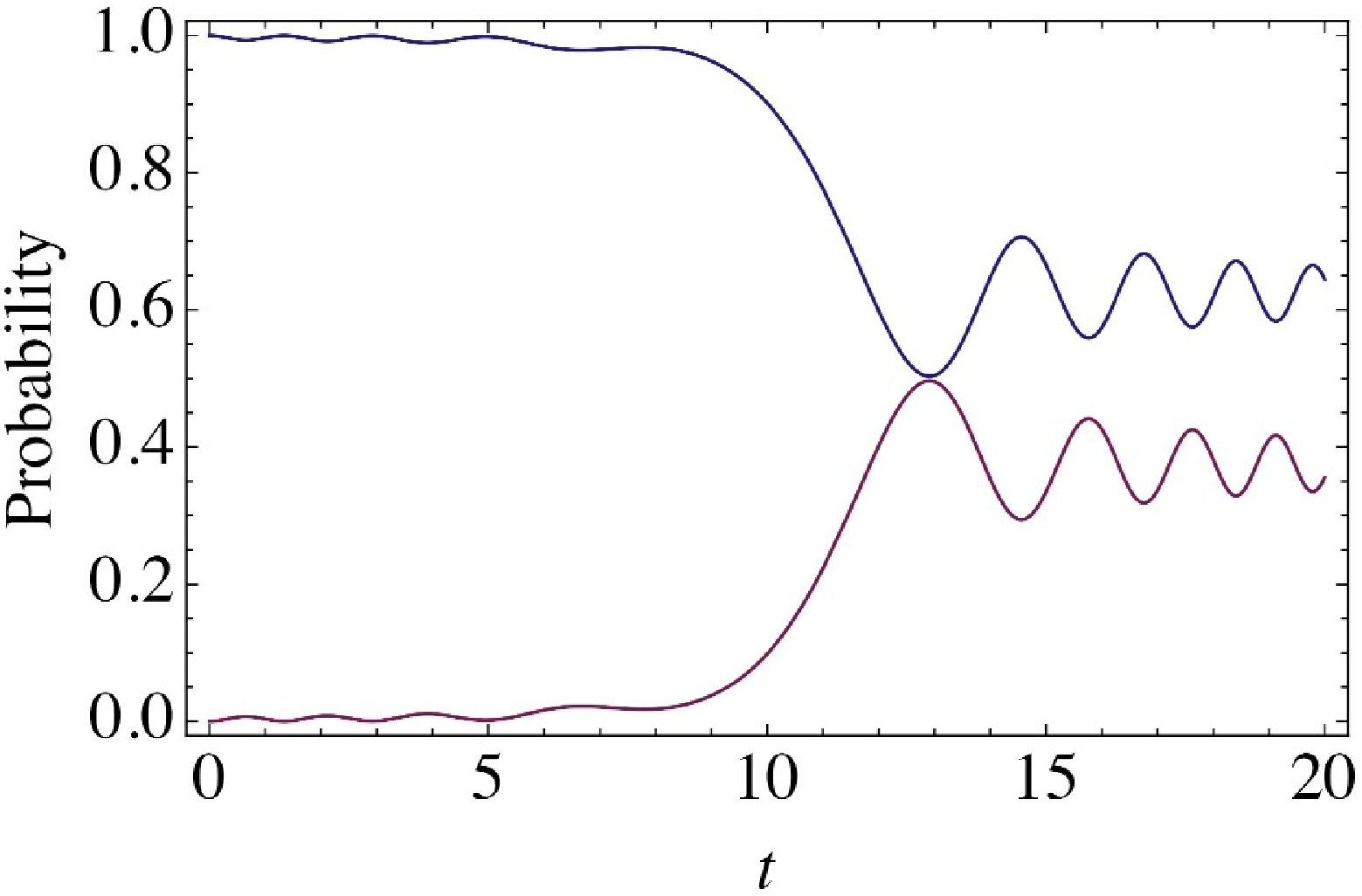}}
\centering\subfigure[]{\includegraphics[width=0.32\textwidth]
{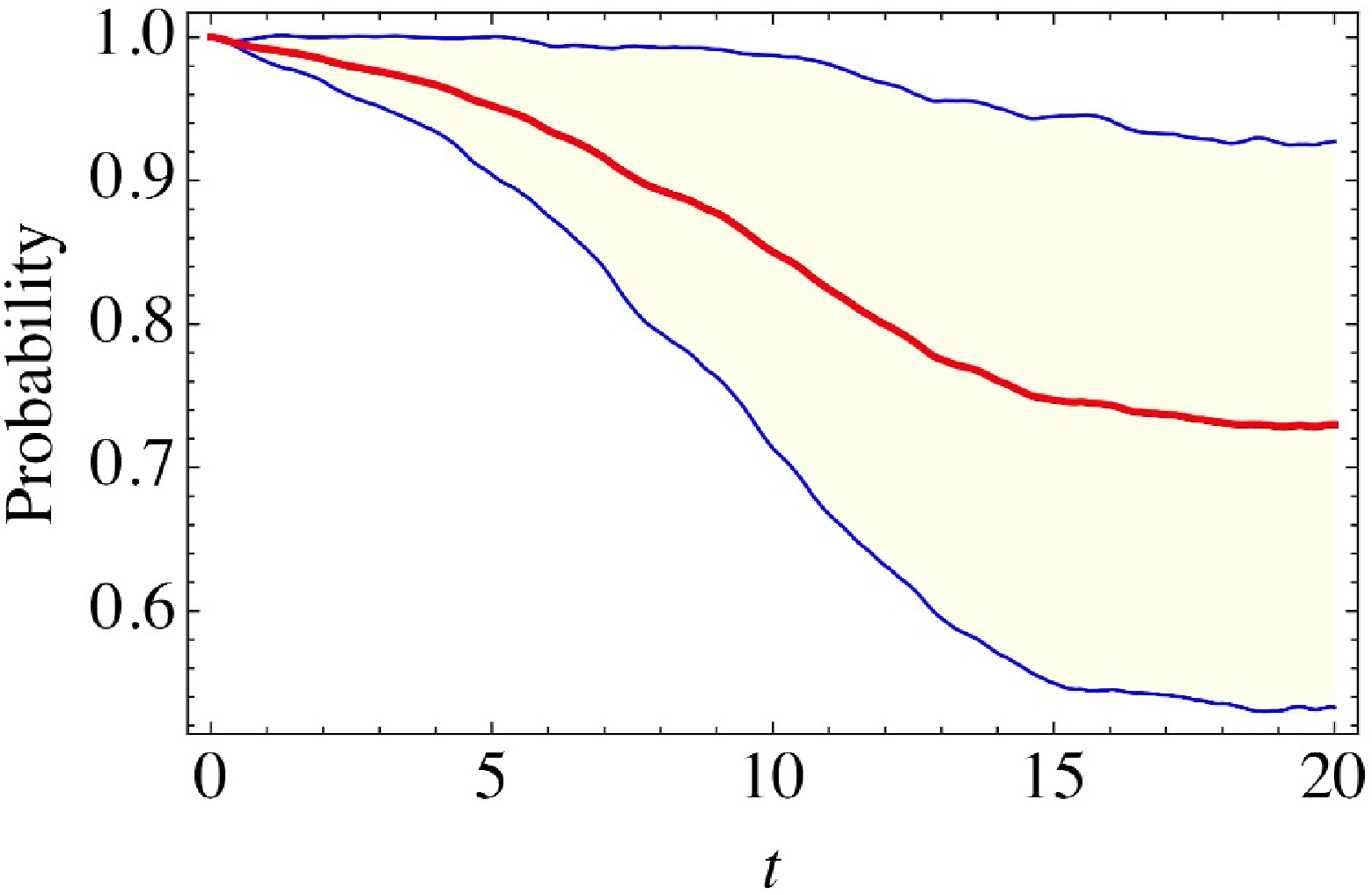}} 
\centering\subfigure[]{\includegraphics[width=0.32\textwidth]
{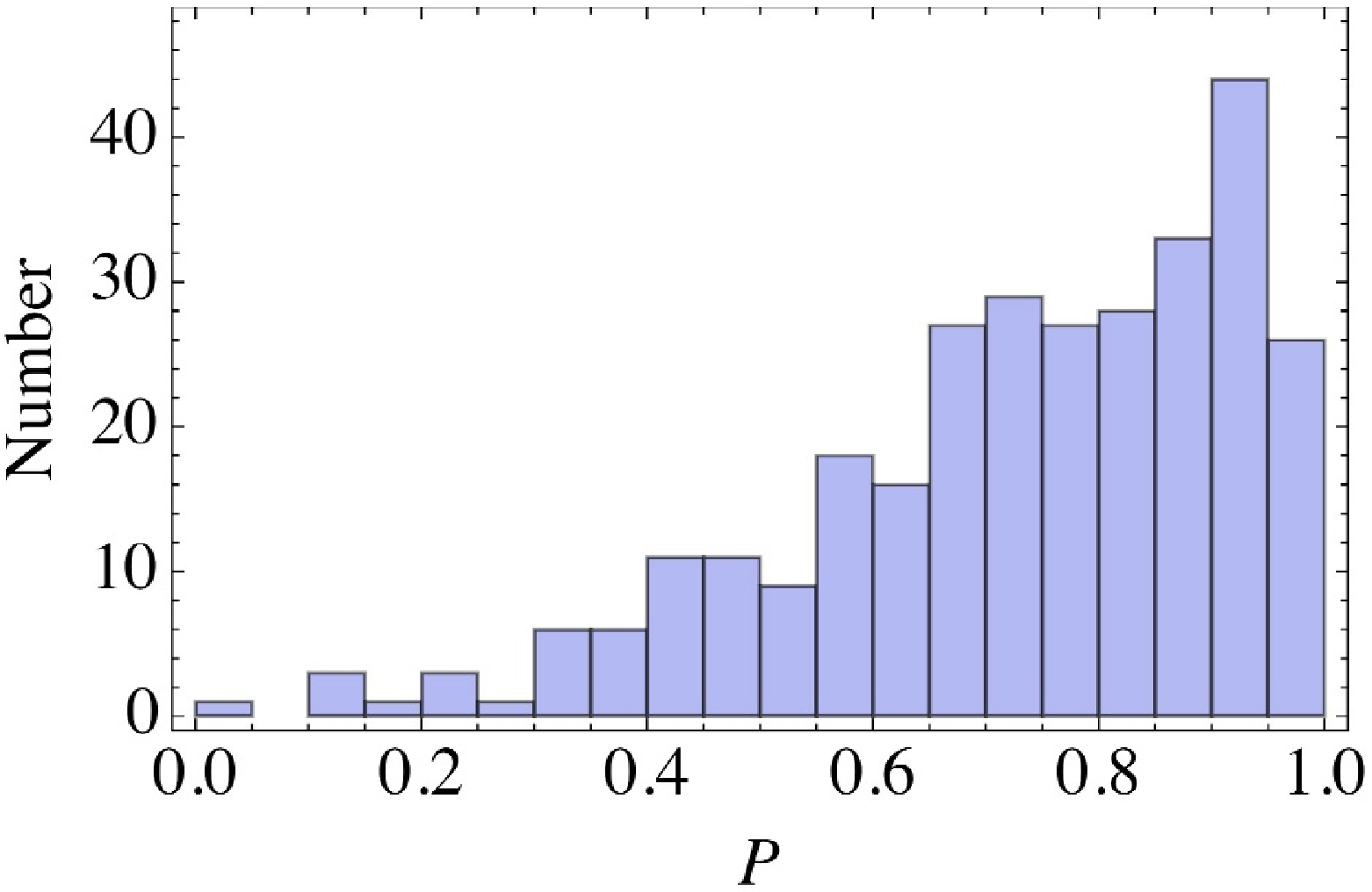}}
\caption{(Color online) (a) Probabilities $P(t;\lambda) \equiv
P_1(t;\lambda) = \psi_1^*(t) \psi_1(t)$ (blue upper curve) and
$P_2(t;\lambda) = \psi_2^*(t) \psi_2(t)$ (red lower curve) versus time
for $\lambda=0.2$ and $T=10$ (the total elapsed time is $2T = 20$)
calculated without dephasing.  (b) Mean and standard deviation of the
LZ probability $P(t;\lambda)$ calculated with dephasing using 300
paths (stochastic realizations) and $\xi_0 = 1$.  (c) Histogram of
$P(t_f;\lambda)$, where $t_f = 2T = 20$, with 300 paths and strong
dephasing, $\xi_0 = 1$.}
\label{Fig_LZ_stochastic_l_0.2_s_1}
\end{figure}

It is instructive to explicitly consider the distributions ${\cal
D}[P(T)]$ for weak and strong couplings.
Figure~\ref{Fig_LZ_stochastic_L_z_Histogram} shows the histogram of
the probabilities at the final time, $P(T) = \psi_1^*(T) \psi_1(T)$
for $\lambda=0.3$, $T=10$ for $\xi_0 = 0.2$ and $\xi_0 = 1.0$.  One
clearly sees that the two distributions are quite different.  For weak
coupling, the distribution is peaked around the mean value which is
the same as for $\xi_0 = 0$, but for strong coupling, the peak of the
distribution is shifted to higher probabilities (near $P=1$) and the
width of the distribution is much broader (standard deviation about
0.3).  This result is in line with the findings in Ref.~\cite{Efrat2}.
Similarly, for $\lambda = 0.2$; at the final time we find,
$\overline{P(\xi_0=1.0)} - P(\xi_0=0) \approx 0.73-0.635$, so the
average probability is shifted to a higher value due to strong
dephasing, and the standard deviation is about 0.2 [see results at the
final time in Fig.~\ref{Fig_LZ_stochastic_l_0.2_s_1}(b)].  Moreover,
the distribution is significantly skewed to higher probabilities (see
below).  Figure~\ref{Fig_LZ_stochastic_l_0.2_s_1}(a) shows the
probabilities $P_1(t)$ and $P_2(t)$ versus time without dephasing for
$\lambda = 0.2$, Fig.~\ref{Fig_LZ_stochastic_l_0.2_s_1}(b) plots the
mean and variance of the LZ probability as a function of time, and
Fig.~\ref{Fig_LZ_stochastic_l_0.2_s_1}(c) shows the histogram of the
probabilities at the final time $P(T)$.  The shifted average
probability and the skewed probability distribution ${\cal D}[P(T)]$
show that dephasing increases the survival probability.  This seems to
be a general trend under relatively strong dephasing conditions (as
opposed to relatively weak dephasing where, even though the standard
deviation of the survival probability can be large, the average
probability is largely unaffected) \cite{comment}.  We note that for
extremely large $\xi_0$, it becomes difficult to control the numerics
so as to maintain unitarity.

\begin{figure} [!hb]
\centering\subfigure[]{\includegraphics[width=0.32\textwidth]
{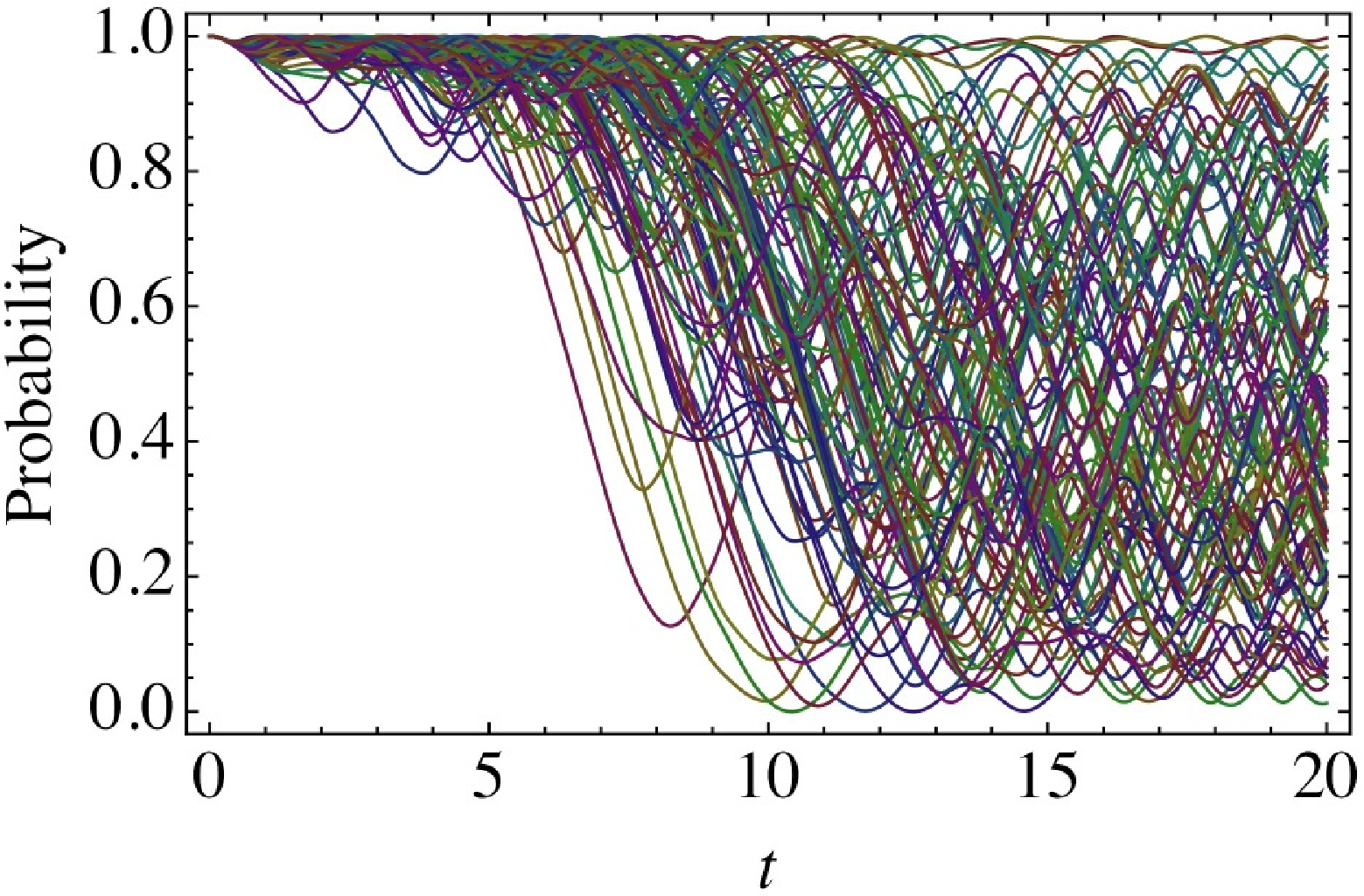}}
\centering\subfigure[]{\includegraphics[width=0.32\textwidth]
{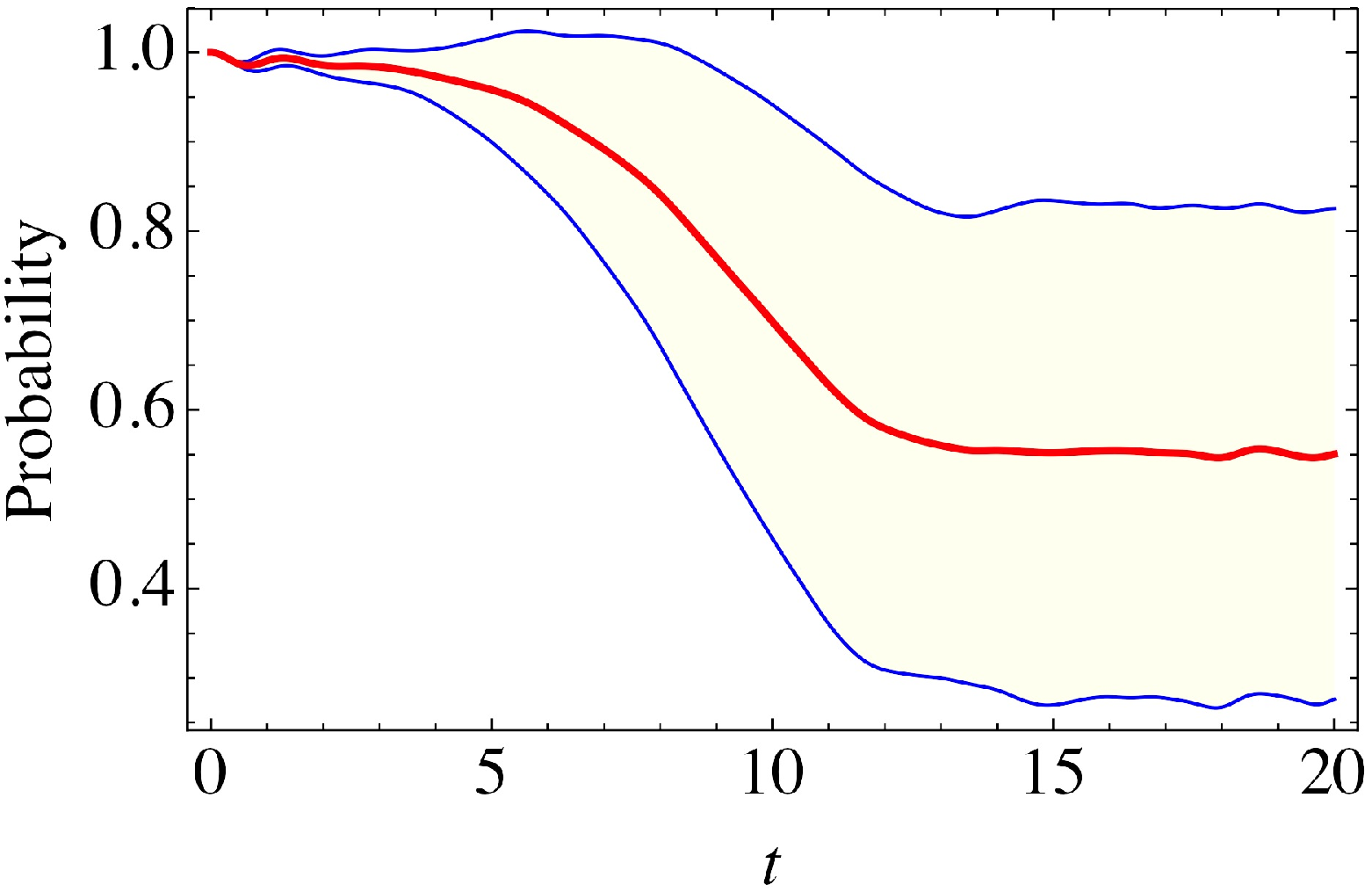}} 
\centering\subfigure[]{\includegraphics[width=0.32\textwidth]
{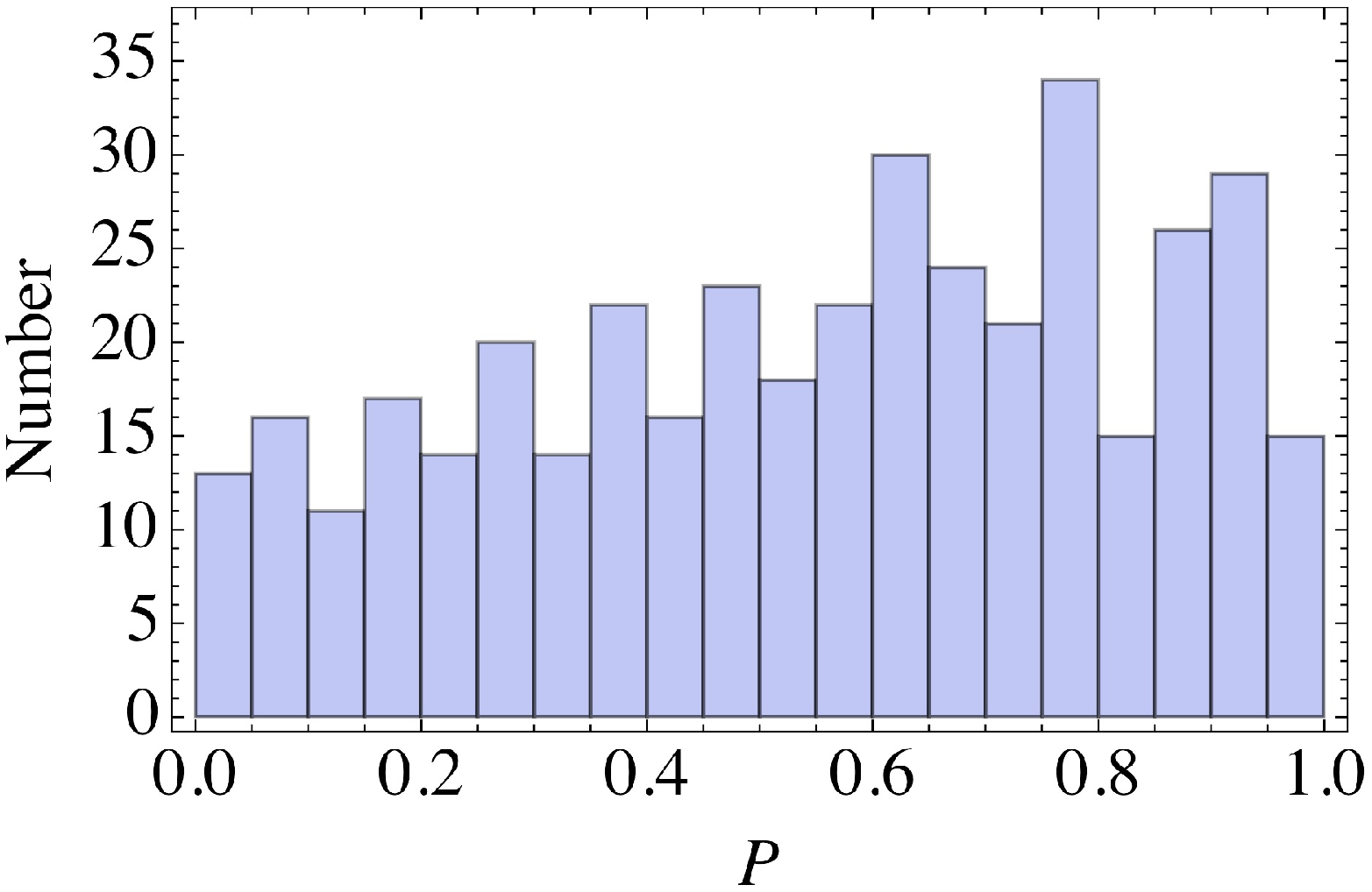}} 
\caption{(Color online) (a) 100 stochastic realizations of the
probability $P(t;\lambda) \equiv P_1(t;\lambda) = \psi_1^*(t)
\psi_1(t)$ for $\lambda = 0.2$ and $T=10$ (the total elapsed time is
$2T = 20$) .  (b) mean and standard deviation of $P(t;\lambda)$ with
400 paths (stochastic realizations) for $\sigma = \vartheta = 1$, $\mu
= 0$ and ${\cal O}_0 = 0$.  (c) Histogram of the probability $P(T;\lambda)$
with 400 paths.}
\label{Fig_LZ_OU_L_z}
\end{figure}

We present results obtained with the same parameters used to obtain
Fig.~\ref{Fig_LZ_stochastic_l_0.2_s_1}, except that now we take the
stochastic process to be the Ornstein--Uhlenbeck process (Brownian
motion).  We use Eqs.~(\ref{SL_stoch_OU}) with $\mu = 0$, $\vartheta =
1$, $\sigma = 1$ and ${\cal O}_0 = 0$.  The dynamics will now not be
Markovian, as opposed to the dyanmics using white Gaussian noise.
Figure \ref{Fig_LZ_OU_L_z}(a) shows 100 stochastic realizations of
$P(t;\lambda)$, Fig.~\ref{Fig_LZ_OU_L_z}(b) shows the mean and
variance of the probability $P(t;\lambda)$ as a function of time
computed with 400 realizations, and \ref{Fig_LZ_OU_L_z}(b) shows the
histogram of the probabilities at the final time, $P(T)$, with 400
realizations.  We see that the mean of the probability goes at the
final time to about 0.55, even for $T = 10$, whereas the white noise
mean is about 0.73 (recall that without dephasing, the final
probability is 0.635 for these conditions).

\begin{figure} [!hb]
\centering\subfigure[]{\includegraphics[width=0.32\textwidth]
{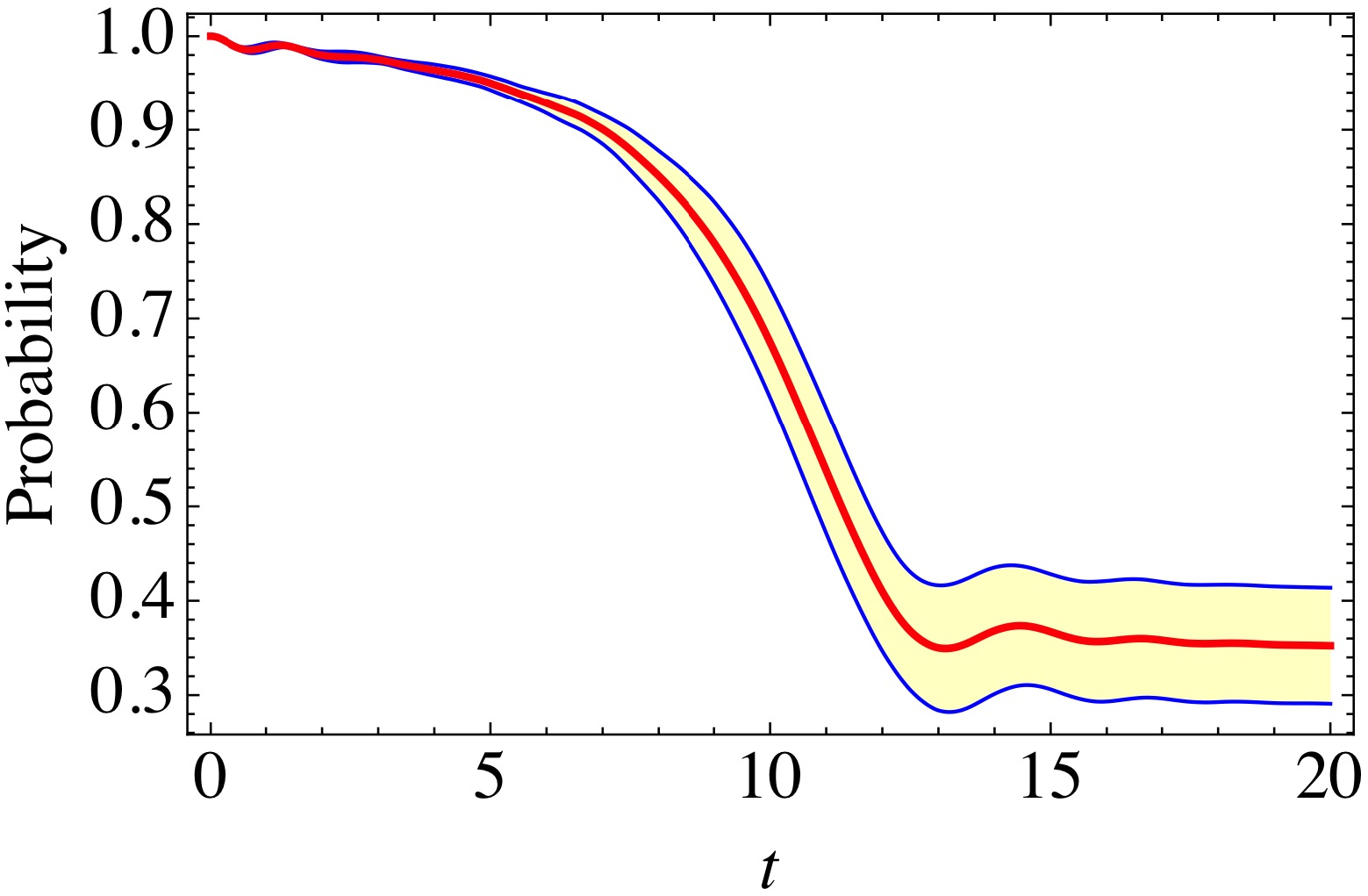}}
\centering\subfigure[]{\includegraphics[width=0.32\textwidth]
{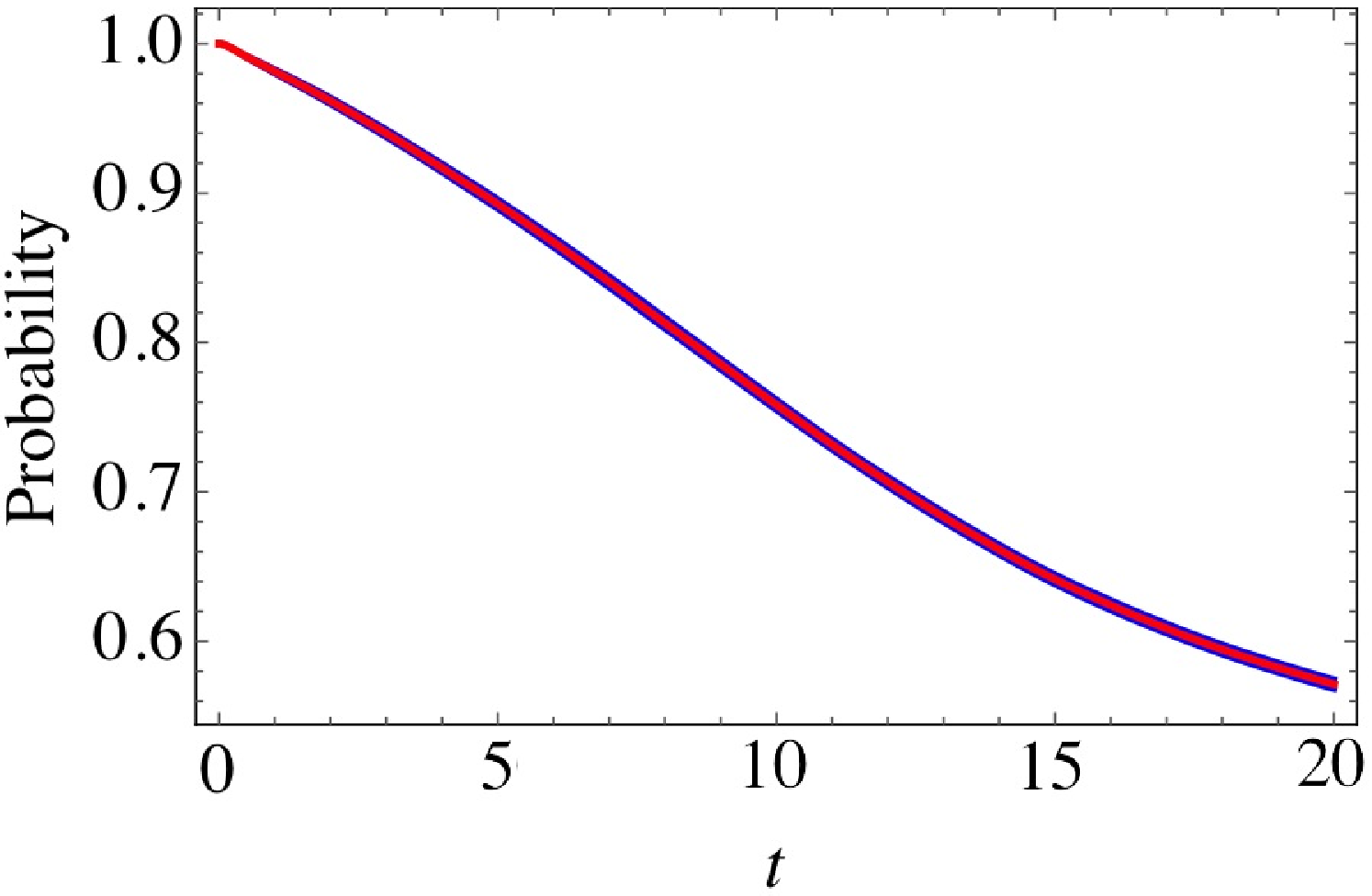}} 
\centering\subfigure[]{\includegraphics[width=0.32\textwidth]
{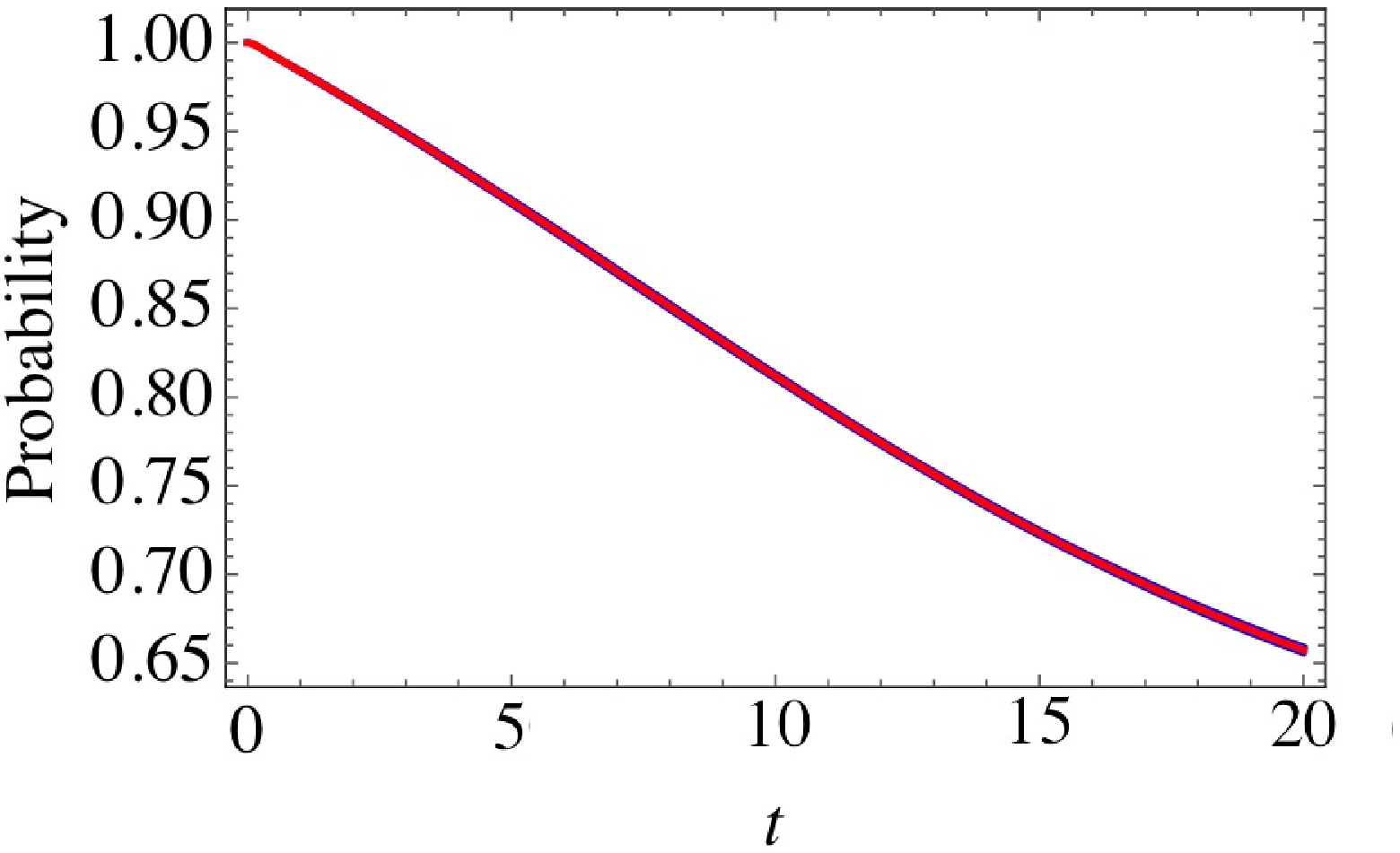}} 
\caption{(Color online) Mean and standard deviation of the probability
$P(t;\lambda,\beta, \xi_0)$ for the LZ problem with one-level decay of strength 
$\beta$, dephasing with sterngth $\xi_0=0.2$, $\lambda = 0.3$, and $T=10$. To be 
compared with Figs.~\ref{LZ_w_decay_beta_10_l_0.3}(b) (decay without dephasing) 
and \ref{Fig_LZ_stochastic_L_z_0.2} (dephasing without decay). 
(a) $\beta=1$ and $\overline{P(T;\lambda,\beta, \xi_0)} \approx 0.3 < \overline{P(T;
\lambda,0, \xi_0)} \approx 0.4$ (see Fig.~\ref{Fig_LZ_stochastic_L_z_0.2}).  (b) $
\beta=10$ and $\overline{P(T;\lambda,\beta, \xi_0)} \approx 0.58 \approx P(T;\lambda,
\beta, 0)$ [see Fig.~\ref{LZ_w_decay_beta_10_l_0.3}(b)]. (c) $\beta=15$ and $
\overline{P(T;\lambda,15, \xi_0)} \approx 0.65 > \overline{P(T;\lambda,10, \xi_0)}$.
The variance in (b) and (c) is very small (barely visible).}
\label{Fig_LZ_stochastic_decay_dephas}
\end{figure}

Finally, we briefly explore the LZ problem with the combined effects of 
one-level decay and dephasing. The object of study is the mean LZ 
probability $\overline{P(T;\lambda,\beta, \xi_0)}$ at long but finite time $T$. The 
questions to be asked are: (1) For fixed decay strength $\beta >0$, 
how does the presence of dephasing $\xi_0 >0$ affect $\overline{P(T;\lambda,\beta, 
\xi_0)}$ as compared with the LZ probability $P(T;\lambda,\beta, 0)$ (no depjasing)? 
(2) For fixed dephasing strength $\xi_0>0$, how does the presence of decay
$\xi_0 >0$ affect the behavior of $\overline{P(T;\lambda,\beta, \xi_0)}$ as function 
of $\beta$?  In other words, is the counterintuitive observation, analyzed previously
in the {\it absence} of dephasing, (that there are situations where $P$ {\it increases} 
with $\beta$), survive also in the presence of dephasing?
To answer these questions we present  the results of calculations based on 
the linear model, with decay and dephasing, i.e., Eqs.~(\ref{SL1_stoch}) and 
(\ref{SL2_stoch}) albeit with $z(t)=-\tfrac{1}{2}(t+i \beta)$, in
Fig.~\ref{Fig_LZ_stochastic_decay_dephas}.  This figure should 
be compared with Fig.~\ref{Fig_LZ_stochastic_L_z_0.2} (which displays the results 
of calculations with dephasing in the absence of decay), and
Fig.~\ref{LZ_w_decay_beta_10_l_0.3} (which displays the results of calculations 
with decay in the absence of dephasing).  The main results of this analysis can be 
stated briefly as follows:  (i) Comparing Figs.~\ref{Fig_LZ_stochastic_decay_dephas}(b) and \ref{LZ_w_decay_beta_10_l_0.3}(b) shows that for large enough $\beta \approx T$, the effect of dephasing on $\overline {P(t;\lambda,\beta, \xi_0)}$ is small. (ii) Comparing the parts of Figs.~\ref{Fig_LZ_stochastic_decay_dephas} with one another and with Fig.~\ref{Fig_LZ_stochastic_L_z_0.2} shows that: (iia) For small $\beta$ the mean probability slightly decreases with increasing decay, but for large $\beta$ the mean probability {\em increases} with increasing $\beta$.  In other words, the answer to question (2) posed above is affirmative. (iib) For large decay rate $\beta \ge T$ the variance of the survival probability shrinks. 

\section{Summary and conclusions}  \label{Sec:Summary}

We studied two aspects of the classical Landau--Zener problem.
\underline {First}, the Landau--Zener problem with decay was analyzed
using a combination of analytic and numeric solutions of the
time-dependent Schr\"odinger equation.  The time dependence of the
energy levels was taken to be either linear $\veps_{1,2}(t)=\pm \alpha
t$ or of the form $\veps_{1,2}(t)=\veps \tanh (t/{\cal T})$.  In the
first case the energies are not bounded as $|t| \to \infty$.  In the
long time limit, the probability $P(\infty)$ is {\em independent} of
decay rate for the linear Landau--Zener case.  This is an artifact of
the unbounded form of the time-dependent energies appearing in the
diagonal elements of the Landau--Zener Hamiltonian.  When the energy
levels are bounded as function of time between the initial and finite
times, the probability {\em does} depend on decay rate.  Surprisingly,
the survival probability of state $\psi_1(t)$ increases with
increasing decay rate $\beta_2$.  This is due to level crossing
(rather than an avoided crossing) that occurs for sufficiently large
$\beta_2$ ($\beta_2 > 4 \lambda$).  These results are valid both for
the linear Landau--Zener problem and for the smoothly saturated
energies of the form $\veps \tanh (t/{\cal T})$.  In the latter case,
the analytic solution for the wave function at large $T$ yields a
particularly simple analytic expression for the probability.

Let us compare our approach with that of Ref.~\cite{Schilling_06},
which is closely related to our study of Landau--Zener with decay.  It
studied the Landau--Zener problem with decay without specifically
specifying the precise time dependence of the two energy levels.
Berry's approach with a superadiabatic basis \cite{Berry_90} is used
to obtain the survival probability $P$ in the slow-sweep limit (small
$\alpha$).  The main result obtained was that $P$ is composed of two
factors, a geometrical and dynamical one, and these factors are
analyzed.  When applied to the model of
Ref.~\cite{Akulin_Schleich_92}, the independence of $P$ on the decay
rate is recovered.  Critical damping of St\"uckelberg oscillations is
predicted and analyzed in the region of very small probability, $P <
10^{-5}$.  Our approach, on the other hand, deals with specific forms
of the time dependence of the energy levels and leads to analytic
solutions of the pertinent second order differential equations.  This
enabled us to carry out a systematic analysis of the dependence of $P$
on decay parameters for arbitrary values of decay rate and channel
coupling.  The independence of $P$ on the decay rate for the linear
case is simply explained in terms of the analytic solution, as is the
dependence of $P$ on the decay rate for the saturated energy case.
Our results are valid for arbitrary sweep rate and interaction
strength, as well as on the decay rates $\beta_1$ and $\beta_2$.

\underline{Second}, we studied a few aspects of the Landau--Zener
problem with dephasing.  For an example of such dephasing processes,
consider the population transfer within the triplet ground state
manifold of diamond NV$^{-}$ centers \cite{Jarmola_12, Doherty_13}.
In diamond NV centers, the $m_s = 0$ level is lower in energy than the
$m_s = \pm 1$ levels due to crystal field effects.  Suppose one is
interested in moving population from $m_s = 0$ to $m_s = -1$ by slowly
sweeping (chirping through resonance) the frequency of a
radio-frequency field that is nearly in resonance with the $m_s = 0
\to -1$ transition.  The $m_s = -1$ state decays to $m_s = 0$
(longitudinal and transverse decay processes can both take place), and
therefore the decay is {\em within} the three-level manifold.  
Generalizing to a stochastic
differential Schr\"{o}dinger--Langevin equation approach enables the
treatment of such cases.  In Sec.~\ref{Sec:dephasing} we carried out
this approach for $T_2$ dephasing during the Landau--Zener dynamics
with both white-noise and Ornstein--Uhlenbeck-noise (a similar
procedure can be used to model $T_1$ processes if the coupling
operator is taken to be ${\cal V} = \sigma_x$ rather than $\sigma_z$).
For Gaussian white noise, his method is equivalent to using a density 
matrix approach with Lindblad operators \cite{vanKampenBook},
but produces non-Markovian dynamics for other kinds of noise.
References~\cite{Rammer} and \cite{Efrat2} showed that for
Landau--Zener transitions with dephasing driven by white noise, the
underlying physics depends on whether the dephasing time is long (weak
dephasing) or short (weak dephasing).  The Schr\"{o}dinger--Langevin
equation approach enabled us to compute the survival probability both
in the weak and strong dephasing regimes.  We calculated the
distribution of the survival probability and pointed out its distinct
behaviour in the long and short dephasing time regimes, and we showed
that Ornstein--Uhlenbeck noise gives somewhat different behavior than
Gaussian white noise. 

We also analyzed the combined effects of one-level decay and dephasing on the 
averaged LZ survival probability. We found that the counterintuitive result, 
that there are situations where the LZ probability  {\em increases} 
with decay rate, survives also in the presence of dephasing.

{\it Acknowledgement.} This work was supported in part by grants from
the Israel Science Foundation (Grant Nos.~400/2012 and 295/2011).  We
are grateful to Professor Dmitry Budker for stimulating our interest
in this problem and for valuable discussions throughout the course of
this work.  Discussions with Efrat Shimshoni and Robert Shekhter are
highly appreciated.



\begin{thebibliography}{99}
\bibitem{comment_AS}
In Ref.~\cite{Akulin_Schleich_92}, the solution of the second order
differential equation (\ref{SL6}) is expressed in terms of a single
transcendental Hermite function, and not as a combination of two
independent transcendental functions, as in Eq.~(\ref{psibc}).  The 
procedure for satisfying the two required initial conditions [the analog of Eq.~(\ref{solC})] is therefore not fully specified.
\bibitem{linear_eig}
These results hold also for the linear Landau--Zener problem with
decay provided $\sinh t$, $\cosh t$ and $\cosh 2t$ are replaced by
$t$, $1$ and $2$ respectively.
\bibitem{comment}
This observation is true for modest $T/\tau_\phi \equiv T \xi_0$, but
when this quantity becomes large, this is no longer true.  For truly
large $\xi_0$ and $T \xi_0$, we expect the survival probability to go
to 1/2.  Moreover, note that it is difficult to control numerics using
the Schr\"{o}dinger-Langevin stochastic differential equation approach
for extremely large $\xi_0$.
\end{thebibliography}
\end{document}